\documentclass[10pt]{article} 
\usepackage{comb, amsmath, latexsym, amssymb, amscd, amsfonts, amsthm}
\usepackage[english]{babel}

\DeclareFixedFont{\sfracFont}{U}{euf}{b}{n}{7pt}
\newtheoremstyle{mydefi}% name
  {15pt}%      Space above
  {15pt}%      Space below
  {}%         Body font
  {}%         Indent amount (empty = no indent, \parindent = para indent)
  {\bfseries}% Thm head font
  {:}%        Punctuation after thm head
  {.5em}%     Space after thm head: " " = normal interword space;
  {}%         Thm head spec (can be left empty, meaning `normal')
\newtheoremstyle{mytheo}% name
  {15pt}%      Space above
  {15pt}%      Space below
  {\slshape}%         Body font
  {}%         Indent amount (empty = no indent, \parindent = para indent)
  {\bfseries}% Thm head font
  {:}%        Punctuation after thm head
  {.5em}%     Space after thm head: " " = normal interword space;
  {}%         Thm head spec (can be left empty, meaning `normal')

\theoremstyle{mytheo}
\newtheorem{stel}{Theorem}[section]
\newtheorem{lem}[stel]{Lemma}

\theoremstyle{mydefi}
\newtheorem{de}[stel]{Definition}

\newcommand{\BB}[1]{\mathbb{#1}}

\newcommand{\ten}{\otimes}
\newcommand{\AC}{\mathcal{A}}
\newcommand{\BC}{\mathcal{B}}
\newcommand{\CC}{\mathcal{C}}
\newcommand{\DC}{\mathcal{D}}
\newcommand{\EC}{\mathcal{E}}
\newcommand{\FC}{\mathcal{F}}
\newcommand{\HC}{\mathcal{H}}
\newcommand{\JC}{\mathcal{J}}
\newcommand{\LC}{\mathcal{L}}
\newcommand{\MC}{\mathcal{M}}

\newcommand{\SC}{\mathcal{S}}
\newcommand{\WC}{\mathcal{W}}
\newcommand{\Id}{\mbox{Id}}

\newcommand{\I}{\mathbf{1}}

\newdimen\mylistindent \mylistindent=20pt  % inspringafstand van \myitem

\numberwithin{equation}{section}
\begin{document}

\title{Stochastic Schr\"odinger equations}
\author{Luc Bouten,~M\u{a}d\u{a}lin Gu\c{t}\u{a}~ and Hans Maassen}
\date{}
\maketitle

\vspace{8mm}

\begin{abstract} 
A derivation of stochastic Schr\"odinger equations is given using quantum filtering
theory. We study an open system in contact with its environment, the electromagnetic 
field. Continuous observation of the field yields information on the system: 
it is possible to keep track in real time of the best estimate of the system's
quantum state given the observations made. This estimate satisfies a stochastic
Schr\"odinger equation, which can be derived from the quantum stochastic 
differential equation for the interaction picture evolution of system and 
field together. Throughout the paper we focus on the basic example of resonance 
fluorescence.
\end{abstract}

\section{Introduction}\label{Introduction}

It has long been recognized that continuous time measurements can not
be described by the standard projection postulate of quantum mechanics.
In the late 60's, beginning 70's, Davies developed a theory
for continuous time measurement \cite{Dav1} culminating in his book 
\cite{Dav}. His mathematical work became known to the quantum optics 
community through the paper with Srinivas on photon 
counting \cite{SrD}.                                      

The late 80's brought renewed interest to the theory of continuous time 
measurement. For instance the waiting time distribution of fluorescence photons 
of a two-level atom driven by a laser was obtained by associating a continuous
evolution to the atom in between photon detections and jumps at the moments
a photon is detected \cite{Car01}. In this way every record of photon detection
times determines a trajectory in the state space of the atom. Averaging
over all possible detection records leads to the well-known description
of the dissipative evolution of the atom by a master equation.
Advantage of the trajectory approach is the fact that an initially 
pure state will remain pure along the whole trajectory. This 
allows for the use of state vectors instead of density matrices, 
significantly speeding up computer simulations \cite{MCD}, \cite{DCM}, 
\cite{GPZ}, \cite{Car}.                                    

Infinitesimally, the quantum trajectories are solutions of a stochastic 
differential equation with the measurement process as the noise term.
The change in the state is given by the sum of two terms: 
a deterministic one proportional with $dt$ and a stochastic one proportional 
to the number of detected photons $dN_t$ in the interval $dt$. 
For other schemes such as homodyne detection the corresponding
stochastic differential equation is obtained as the diffusive limit
of photon counting where the jumps in the state space 
decrease in size but become increasingly frequent 
\cite{BaB}, \cite{Car}, \cite{WiM}. In this limit the stochastic term 
in the differential equation is replaced by a process with continuous
paths.      

The stochastic Schr\"odinger equations obtained in this way had been 
postulated before by Gisin \cite{Gis}, \cite{Gis1}, \cite{DGHP},
in an attempt to generalize the customary unitary evolution in 
quantum mechanics. The stochastic terms are seen as randomness 
originating from the measurement process. However, in this approach 
the correspondence between the different quantum state diffusion equations 
and the measurements that can be performed is not emphasized.  

Another approach originated from the development of quantum stochastic
calculus \cite{HuP}, \cite{Par}, generalizing the classical It\^o table to quantum 
noises represented by creation and annihilation operators (see Section \ref{QSC}).
Barchielli saw the relevance of this new calculus for quantum optics \cite{Bar}. 
Indeed, in the Markovian approximation the interaction between a quantum system
and the electromagnetic field is governed by a unitary solution of a quantum 
stochastic differential equation in the sense of \cite{HuP}.

Belavkin was the first to see the connection between quantum measurement 
theory and classical filtering theory \cite{Kal}, in which one estimates 
a signal or system process when observing a function of the signal in the 
presence of noise. This is done by deriving the filtering equation
which is a stochastic differential equation for the expectation value of the system 
process conditioned on outcomes of the observation process. Belavkin extended
the filtering theory \cite{Bel}, \cite{Bel01} to allow for the quantum noises
of \cite{HuP}. Stochastic Schr\"odinger equations turn out to be examples of the 
quantum filtering or Belavkin equation \cite{Bel0}, \cite{BeS}.       

Aim of this paper is to give an elementary presentation of quantum filtering
theory. We construct the expectation of an observable conditioned on outcomes
of a given measurement process. The differential form of this conditional
expectation is the stochastic Schr\"odinger equation associated with the 
given measurement. At the heart of the derivation lies 
the It\^o table of quantum stochastic calculus enabling a fast computation 
of the equation. The procedure is summarized in a small recipe in Section
\ref{Belavkinslemma}.       

To illustrate the theory we consequently focus on the basic example of 
resonance fluorescence of a two-level atom for which we consider 
photon counting and homodyne detection measurement schemes. The stochastic
Schr\"odinger equations for these examples are derived in two ways, once
via the usual approach using quantum trajectories and a diffusive limit, and
once using quantum filtering theory. In this way we hope to emphasize how 
conceptually different both methods are. 

This paper is organised as follows. Sections \ref{Daviesprocess} and 
\ref{Homodynedetection} serve as an introduction to the guiding example
of this paper: resonance fluorescence of a two-level atom driven by a laser. 
In Section \ref{Daviesprocess} we put the photon counting description of 
resonance fluorescence by Davies \cite{BMK}, \cite{Car01}, \cite{Car0} into 
the form of a stochastic differential equation driven by the counting process. 
In Section \ref{Homodynedetection} we discuss the 
homodyne detection scheme as a diffusive limit of the photon counting 
measurement, arriving at a stochastic differential equation driven by a 
diffusion process. The equations of Sections \ref{Daviesprocess}
and \ref{Homodynedetection} will be rederived later in a more general way 
using quantum filtering theory. 

In Section \ref{conditional expectation} we introduce the concept of
conditional expectation in quantum mechanics by first illustrating it 
in some simple, motivating examples. Section \ref{dilation}  
describes the dissipative evolution of the open system 
within the Markov approximation. The joint evolution of the 
system and its environment, the quantized electromagnetic field, is 
given by unitaries satisfying a quantum stochastic differential equation. 
Given a measurement of some field observables it is shown how to 
condition the state of the system on outcomes of the measurement 
using the construction of Section \ref{conditional expectation}.
Section \ref{QSC} is a short review of quantum stochastic
calculus and its applications to open systems. Sections \ref{dilation}
and \ref{QSC} describe dilation theory and quantum stochastic calculus
in a nutshell. 

Section \ref{Belavkinslemma} contains the derivation of the quantum 
filtering equation, the stochastic differential equation for the 
conditional expectation. This equation is the stochastic Schr\"odinger 
equation for the given measurement. This part ends with a recipe
for computing stochastic Schr\"odinger equations for a large class
of quantum systems and measurements. The end of the article
connects to Sections \ref{Daviesprocess} and \ref{Homodynedetection}
by showing how the recipe works in our main example.

\section{The Davies process}\label{Daviesprocess}

We consider a two-level atom in interaction with the quantized electromagnetic field. 
The state of the atom is described by a $2 \times 2$-density matrix $\rho$, i.e.\ 
$\rho\geq 0$, and $\mbox{Tr}\rho=1$. Atom and field together perform a unitary, thus 
reversible evolution, but by taking a partial trace over the electromagnetic field we 
are left with an irreversible, dissipative evolution of the atom alone. 
In the so called Markov limit it is given by a norm continuous semigroup 
$\{T_t\}_{t \ge 0}$ of completely positive maps. A central example discussed in this paper 
is resonance fluorescence. Here the atom is driven by a laser on the {\it forward} channel, while 
in the {\it side} channel a photon counting measurement is performed. For the time being we will 
suppress the oscillations of the laser for reasons of simplicity. 
In this case the Lindblad generator of $T_t$, or Liouvillian $L$ is given by (cf.\ \cite{Car}):
  \begin{equation}\label{Master}
  \frac{d}{dt}\Big|_{t=0}T_t(\rho) = 
  L(\rho) = -i[H, \rho] + i\frac{\Omega}{2}[V+V^*, \rho] - \frac{1}{2}\{V^*V,\rho\} + V \rho V^*, 
\mbox{\ where \ }
  V = \begin{pmatrix} 0 & 0 \\ 1 & 0 \end{pmatrix},
  \end{equation}
$H := \frac{\omega_0}{2}\sigma_z$ is the Hamiltonian of the atom, and $\Omega$ is the 
\emph{Rabi frequency}.  \\
The master equation \eqref{Master} can be \emph{unravelled} in many ways 
depending on what photon detection measurement is performed. By unravelling the master equation 
we mean writing $L$ as the sum $\LC +\JC$, where $\JC$ represents the instantaneous 
state change taking place when detecting a photon, and $\LC$ describes the smooth state 
variation in between these instants. The unravelling for photon counting in the side
channel is given by \cite{Car}    
  \begin{equation*}
  \LC(\rho) =-i[H, \rho] + i \frac{\Omega}{2}[V+V^*,\rho] - \frac{1}{2}\{V^*V,\rho\} + 
(1-|\kappa_s|^2)V\rho V^* \mbox{\ and \ } \JC(\rho) = |\kappa_s|^2 V\rho V^*,    
  \end{equation*}
with $|\kappa_s|^2$ the decay rate into the side channel.        \\ 
An outcome of the measurement over an arbitrary finite time interval $[0,t)$
is the set of times $\{t_1, t_2, \ldots, t_k\}$ at which photons are detected 
in the side channel of the field. The number of detected photons can be arbitrary, thus the 
space of outcomes is 
 \begin{equation*}
 \Omega\left([0,t)\right) := \bigcup_{n=0}^\infty \Omega_n\left([0,t)\right)=
 \bigcup_{n=0}^\infty  \{\sigma \subset [0,t);\ |\sigma| = n\}
 \end{equation*}
also called the \emph{Guichardet space} \cite{Gui}. 
In order to describe the probability distribution of the 
outcomes we need to make $\Omega\left([0,t)\right)$ into a measure space. 
Let us consider the space of $n$-tuples $[0,t)^n$ with its Borel 
$\sigma$-algebra and the measure $\frac{1}{n!}\lambda_n$ where $\lambda_n$ is the 
Lebesgue measure. Then the map
 \begin{equation*} 
 j_n: [0,t)^n\ni (t_1,\dots, t_n)\to\{t_1,\dots, t_n\} \in \Omega_n\left([0,t)\right)
 \end{equation*}
induces  the $\sigma$-algebra $\Sigma_n\left([0,t)\right)$ and the measure $\mu_n$ on 
$\Omega_n\left([0,t)\right)$.  We define now the measure 
$\mu$ on $\Omega\left([0,t)\right)$ such that $\mu(\{\emptyset\})= 1$ and $\mu = \mu_n$ on 
$\Omega_n\left([0,t)\right)$. 
We will abbreviate $\Omega\big([0,t)\big)$ and $\Sigma\big([0,t)\big)$ to $\Omega^t$ and 
$\Sigma^t$, respectively.                              \\
Davies was the first to show \cite{Dav} 
(see also \cite{Car}, \cite{BMK}) that the unnormalized state of the 
$2$-level atom at time $t$ with initial state $\rho$, and 
conditioned on the outcome of the experiment being in a set $E\in\Sigma^t$ is given by: 
  \begin{equation*}
  \MC^t[E](\rho) = \int_E W_{t}(\omega)(\rho) d\mu(\omega),
  \end{equation*} 
where for $\omega = \{t_1, \ldots, t_k\} \in \Omega^t$ with 
$0 \le t_1 \le \ldots \le t_k < t$ we have
  \begin{equation*}
  W_{t}(\omega)(\rho) := \exp\big((t-t_k)\LC\big)\JC \ldots \JC\exp\big((t_2-t_1)\LC\big)\JC
  \exp\big(t_1\LC\big)(\rho).
  \end{equation*}
Furthermore, $\BB{P}_\rho^t[E] :=\mbox{Tr}(\MC^t[E](\rho))$ is the probability that the event 
$E$ occurs if the initial state is $\rho$. 
% In this way we get a family of probability measures 
% $\{\BB{P}_\rho^t\}_{t \ge 0}$ defined
% by $\BB{P}_\rho^t[E] := \mbox{Tr}\big(\MC^t[E](\rho)\big)$ on the sigma fields 
% $\Sigma^t (t\ge 0)$. 
The family of prabability measures $\{\BB{P}_\rho^t\}_{t \ge 0}$ is consistent, 
i.e.\ $\BB{P}_\rho^{t+s}[E] = \BB{P}_\rho^t[E]$ for all $E \in \Sigma^t, s\ge 0$, 
see \cite{BMK}, hence by Kolmogorov's extension theorem it extends to a single 
probability measure $\BB{P}_\rho$ on
the $\sigma$-algebra $\Sigma^\infty$, of the set $\Omega^\infty$. \\
On the measure space $(\Omega^\infty, \Sigma^\infty, \BB{P}_\rho)$ we define the following random 
variables:
  \begin{equation*}
  N_t:\ \Omega^\infty \to \BB{N}:\ \omega \mapsto |\omega \cap [0,t)|,
  \end{equation*}    
counting the number of photons detected in the side channel up to time $t$. 
The counting  process $\{N_t\}_{t\geq 0}$ has differential $dN_t := N_{t+dt}-N_t$ satisfying 
$dN_t(\omega) = 1$ if $t\in \omega$ and $dN_t(\omega) = 0$ otherwise. 
Therefore we have the following It\^o rules: $dN_tdN_t = dN_t$ and $dN_tdt = 0$, 
(cf.\ \cite{BaB}). \\
To emphasise the fact that the evolution of the $2$-level atom is stochastic, we will regard the 
normalized density matrix as a random variable $\{\rho^t_\bullet\}_{t \ge 0}$ 
with values in the $2 \times 2$-density matrices defined as follows:
  \begin{equation}\label{rhoomega}
  \rho^t_\bullet:\ \Omega^\infty \to M_2:\ \omega \mapsto \rho^t_\omega := 
  \frac{W_{t}\big(\omega\cap [0,t)\big)(\rho)}
  {\mbox{Tr}\Big(W_{t}\big(\omega\cap [0,t)\big)(\rho)\Big)}.
  \end{equation}
The processes $N_t$ and $\rho^t_\bullet$ are related through the stochastic differential 
equation $d\rho^t_\bullet = \alpha_t dt + \beta_t dN_t$. Following \cite{BaB} we will 
now determine the processes $\alpha_t$ and $\beta_t$ by differentiating \eqref{rhoomega}. 
If $t \in \omega$ then $dN_t(\omega) = 1$, i.e.\ the differential
$dt$ is negligible compared to $dN_t=1$, therefore:
  \begin{equation}\label{beta}
  \beta_t(\omega) = \rho^{t+dt}_\omega - \rho^t_\omega = \frac{\JC(\rho^t_\omega)}
  {\mbox{Tr}\big(\JC(\rho^t_\omega)\big)} - \rho^t_\omega.
  \end{equation}
On the other hand, if $t \not \in \omega$ then $dN_t(\omega) = 0$, i.e.\ $dN_t$ is 
negligible compared to $dt$. Therefore it is only the $dt$ term that contributes:
  \begin{equation}\label{alpha}\begin{split}
  &\alpha_t(\omega) = \frac{d}{ds}\Big|_{s=t} \frac{\exp\big((s-t)\LC\big)(\rho^t_\omega)}
  {\mbox{Tr}\Big(\exp\big((s-t)\LC\big)(\rho^t_\omega)\Big)} =  \\
  &\LC(\rho^t_\omega) - \frac{\rho^t_\omega}{\mbox{Tr}(\rho^t_\omega)^2}
\mbox{Tr}\big(\LC(\rho^t_\omega)\big) 
  = \LC(\rho^t_\omega) + \mbox{Tr}\big(\JC(\rho^t_\omega)\big)\rho^t_\omega,
  \end{split}\end{equation}    
where we used that $\mbox{Tr}\big(\LC(\rho^t_\omega)\big) = - 
\mbox{Tr}\big(\JC(\rho^t_\omega)\big)$, 
as a consequence of the 
fact that \ $\mbox{Tr}\big(L(\sigma)\big) = 0$ for all density matrices $\sigma$.
Substituting \eqref{beta} and \eqref{alpha} into 
$d\rho^t_\bullet = \alpha_t dt  + \beta_t dN_t$ we 
get the following \emph{stochastic Schr\"odinger equation} for the
state evolution of the $2$-level atom if we are counting photons in the side channel 
(cf.\ \cite{BaB}, \cite{Car0}):
  \begin{equation}\label{Belcount}
  d\rho^t_\bullet = L(\rho^t_\bullet)dt + \Big(\frac{\JC(\rho^t_\bullet)}
 {\mbox{Tr}\big(\JC(\rho^t_\bullet)\big)}- \rho^t_\bullet\Big)\Big
 (dN_t - \mbox{Tr}\big(\JC(\rho^t_\bullet)\big)dt\Big).
  \end{equation}
The differential $dM_t := dN_t - \mbox{Tr}\big(\JC(\rho^t_\bullet)\big)dt$ and the initial 
condition $M_0 = 0$ define an important process $M_t$ called the \emph{innovating martingale}, 
discussed in more detail in Section \ref{Belavkinslemma}.

\section{Homodyne detection}\label{Homodynedetection}

We change the experimental setup described in the previous section by introducing a 
\emph{local oscillator}, i.e.\ a one mode oscillator in a coherent state given
by the normalised vector in $l^2(\BB{N})$  
 \begin{equation}\label{eq.coherent}
 \psi(\alpha_t) :=\exp\big(\frac{-|\alpha_t|^2}{2}\big)(1, \alpha_t, \frac{\alpha_t^2}{\sqrt{2}}, 
 \frac{\alpha_t^3}{\sqrt{6}}, \ldots),
 \end{equation} 
for a certain $\alpha_t \in \BB{C}$. We take $\alpha_t = \frac{w_t}{\varepsilon}$, 
where $w_t$ is a complex number with modulus $|w_t| = 1$, and  $\varepsilon > 0$. 
The number $\varepsilon$ is inversely proportional to the intensity 
of the oscillator. Later on we will let the intensity go to infinity, i.e.\ 
$\varepsilon\to 0$. The phase $\phi_t$ of the oscillator is represented by 
$w_t = \exp(i\phi_t)$, with $\phi_t = \phi_0 + \omega_{lo} t$, where $\omega_{lo}$ 
is the frequency of the oscillator.                 \\
The local oscillator is coupled to a channel in the electromagnetic field, 
the local oscillator beam. The field is initially in the vacuum state. The 
local oscillator and the field are coupled in such a way that every time a photon is detected 
in the beam, a jump on the local oscillator occurs, given by the operation  
  \begin{equation}
  \JC_{lo}(\rho) = A_{lo}\rho A^*_{lo},
  \end{equation} 
where $A_{lo}$ is the annihilation operator corresponding to the mode of the local 
oscillator. The coherent state $\psi(\alpha_t)$ is an eigenstate of the jump operator $A_{lo}$ at 
eigenvalue $\alpha_t$.  \\
Now we are ready to discuss the homodyne detection scheme. Instead of directly counting photons
in the side channel we first mix the side channel with the local oscillator beam 
with the help of a fifty-fifty beam splitter. In one of the emerging beams a 
photon counting measurement is performed. A detected photon can come from the atom through the side 
channel or from the local oscillator via the local oscillator beam. 
Therefore the jump operator on states $\sigma$ of the atom and the oscillator together, 
is the sum of the respective jump operators:
  \begin{equation*}
  \JC_{a\ten lo}(\sigma) =   
  (\kappa_s V \ten I + I \ten A_{lo})\sigma
  (\overline{\kappa}_s V^* \ten I + I \ten A^*_{lo}).  
  \end{equation*}   
An initial product state $\rho \ten |\psi(\alpha_t)\rangle\langle\psi(\alpha_t)|$ of 
the $2$-level atom and the local oscillator will remain a product after the jump 
since $\psi(\alpha_t)$  is an eigenvector of the annihilation operator. Tracing out 
the local oscillator yields the following jump operation for the atom in the homodyne 
setup: 
  \begin{equation*}\begin{split}
  \JC_a(\rho) = 
  \mbox{Tr}_{lo} \Big(\JC_{a\ten lo}\big(\rho\ten \big|\psi(\alpha_t)\big\rangle
  \big\langle\psi(\alpha_t)\big|\big)\Big) 
  = \big(\kappa_s V + \frac{w_t}{\varepsilon}\big)
  \rho\big(\overline{\kappa}_s V^* + \frac{\overline{w}_t}{\varepsilon}\big).
  \end{split}\end{equation*}
In the same way as in Section \ref{Daviesprocess}, we can derive the following 
stochastic Schr\"odinger equation for the state evolution of the two-level atom when
counting photons after mixing the side channel and the local oscillator beam \cite{BaB} \cite{Car0}:
  \begin{equation}\label{Belcounthomodyne}
  d\rho^t_\bullet = L(\rho^t_\bullet)dt + \frac{1}{\varepsilon}\Big(\frac{\JC_a(\rho^t_\bullet)}
 {\mbox{Tr}\big(\JC_a(\rho^t_\bullet)\big)}- \rho^t_\bullet\Big)
 \varepsilon\Big(dN_t - \mbox{Tr}\big(\JC_a(\rho^t_\bullet)\big)dt\Big),
  \end{equation}   
where the extra $\varepsilon$'s  are introduced for future convenience. 
We will again use the abbreviation: $dM^a_t = dN_t - \mbox{Tr}\big(\JC_a(\rho^t_\bullet)\big)dt$ 
for the innovating martingale (see Section \ref{Belavkinslemma}).
In the homodyne detection scheme the intensity of the local oscillator beam is 
taken extremely large, i.e.\ we are interested in the limit $\varepsilon \to 0$ \cite{BaB}, 
\cite{Car}, \cite{WiM}. Then the number of detected photons becomes very large and it makes sense 
to scale and center $N_t$, obtaining in this way the process with differential 
$dW_t^\varepsilon := \varepsilon dN_t -dt/\varepsilon$ and $W_0^\varepsilon = 0$. 
We find the following It\^o rules for $dW_t^\varepsilon$:
  \begin{equation*}\begin{split}
  &dW^\varepsilon_tdW^\varepsilon_t = \big(\varepsilon dN_t - \frac{1}{\varepsilon}dt\big)
    \big(\varepsilon dN_t - \frac{1}{\varepsilon}dt\big) = \varepsilon^2 dN_t = 
    \varepsilon dW^\varepsilon_t + dt,                                                                 \\
  &dW^\varepsilon_t dt = 0.                                                    
  \end{split}\end{equation*}
In the limit $\varepsilon \to 0$ this becomes $dW_tdW_t = dt$ and $dW_tdt = 0$, i.e.\
the process $W_t := \lim_{\varepsilon \to 0} W_t^\varepsilon$ is a diffusion. 
It is actually this scaled and centered process that is being observed and not 
the individual photon counts $N_t$, see \cite{Car}. 
We pass now to the evaluation of the limit of \eqref{Belcounthomodyne}:
  \begin{equation*}
  \lim_{\varepsilon \to 0}\frac{1}{\varepsilon}\Big(\frac{\JC_a(\rho^t_\bullet)}
 {\mbox{Tr}\big(\JC_a(\rho^t_\bullet)\big)}- \rho^t_\bullet\Big) = w_t \overline{\kappa}_s
  \rho^t_\bullet V^* + \overline{w}_t\kappa_s V \rho^t_\bullet - \mbox{Tr}(w_t \overline{\kappa}_s
  \rho^t_\bullet V^* + \overline{w}_t\kappa_s V \rho^t_\bullet)\rho^t_\bullet.
  \end{equation*}
This leads to the following stochastic Schr\"odinger equation for the
homodyne detection scheme \cite{BaB}, \cite{Car0}, \cite{WiM}
  \begin{equation}\label{Belhomodyne}
  d\rho^t_\bullet = L(\rho^t_\bullet)dt + \left(w_t \overline{\kappa}_s
  \rho^t_\bullet V^* + \overline{w}_t\kappa_s V \rho^t_\bullet - \mbox{Tr}(w_t \overline{\kappa}_s
  \rho^t_\bullet V^* + \overline{w}_t\kappa_s V \rho^t_\bullet)\rho^t_\bullet~\right)dM^{hd}_t, 
  \end{equation}
 for all states $\rho \in M_2$, where 
  \begin{equation}
  dM_t^{hd} := dW_t - \mbox{Tr}(w_t \overline{\kappa}_s
  \rho^t_\bullet V^* + \overline{w}_t\kappa_s V \rho^t_\bullet)dt. 
  \end{equation} 

Let $a_s(t)$ and $a_b(t)$ denote the annihilation operators for the side channel
and the local oscillator beam, respectively. They satisfy the canonical commutation relations
  \begin{equation*}
  [a_i(t), a^*_j(r)] = \delta_{i,j} \delta(t-r), \ \ \ i,j \in \{s,b\}. 
  \end{equation*}
Smearing with a quadratically integrable function $f$ gives 
  \begin{equation*}
  A_i(f) = \int f(t)a_i(t)dt, \ \ \ i \in \{s,b\}.
  \end{equation*}  
By definition, the stochastic process $\{N_t\}_{t \ge 0}$ 
counting the number of detected photons has the same law as the the number operator $\Lambda(t)$ 
up to time $t$ for the beam on which the measurement is performed. Formally we can write
   \begin{equation*}\begin{split}
  \Lambda(t) = 
  \int_0^t\big(a^*_s(r) \ten I + I \ten a^*_b(r)\big)
    \big(a_s(r) \ten I + I \ten a_b(r)\big)dr.   
  \end{split}\end{equation*}
The oscillator beam is at time $t$ in the coherent state 
$\psi\left(\frac{f_t}{\varepsilon}\right)$, where $f_t \in L^2(\BB{R})$ is the function 
$r \mapsto w_r\chi_{[0,t]}(r)$.  
Since the state of the local oscillator beam is an eigenvector of the annihilation operator 
$a_b(r)$ 
  \begin{equation*}
  a_b(r)\psi\left(\frac{f_t}{\varepsilon}\right)=\frac{w_r}{\varepsilon}
  \psi\left(\frac{f_t}{\varepsilon}\right),
  \end{equation*}
we find 
 \begin{equation*}\begin{split}
 \varepsilon \Lambda(t)-\frac{t}{\varepsilon} & =
   \varepsilon \Lambda_s(t)\ten I + 
   \varepsilon \int_0^t \big(\frac{w_r}{\varepsilon}a^*_s(r) + 
  \frac{\overline{w}_r}{\varepsilon}
  a_s(r)\big)\ten I + \frac{|w_r|^2}{\varepsilon^2}dr - \frac{t}{\varepsilon} \\
 & = \varepsilon \Lambda_s(t)\ten I + \big(A^*_s(f_t) + A_s(f_t)\big)\ten I.
 \end{split}\end{equation*}
The operator $X_\phi(t) :=  A^*_s(f_t) + A_s(f_t)$ is called a {\it field quadrature}.
We conclude that in the limit $\varepsilon\to 0$ the homodyne detection is a setup for 
continuous time measurement of the field quadratures $X_\phi(t)$ of the side channel.
(cf.\ \cite{Car}).

\section{Conditional expectations}\label{conditional expectation}

In the remainder of this article we will derive the equations \eqref{Belcount} and 
\eqref{Belhomodyne} in a different way. We will develop a general way to derive Belavkin 
equations (or stochastic Schr\"odinger equations). 
The counting experiment and the homodyne detection experiment, described in the 
previous sections, serve as examples in this general framework. 
The method we describe here closely
follows Belavkin's original paper on quantum filtering theory \cite{Bel}. Our approach differs in 
its construction of the conditional expectation, which is the topic of this section.                  \\
Let us remind the concept of conditional expectation from probability theory. 
Let $(\Omega, \Sigma, \mathbb{P})$ be a probability space describing the ``world'' 
and $\Sigma'\subset\Sigma$ a $\sigma$-algebra of events to which ``we have access''. 
A random variable $f$ on $(\Omega, \Sigma, \mathbb{P})$ with $\mathbb{E}(|f|)<\infty$ can 
be projected to its conditional expectation 
$\mathcal{E}(f)$ which is measurable with respect to $\Sigma'$ and satisfies 
 \begin{equation*} 
 \int_{E}f\mbox{d}\mathbb{P}= \int_{E}\mathcal{E}(f)\mbox{d}\mathbb{P} 
 \end{equation*}
for all events $E$ in $\Sigma'$. Our information about the state of that 
part of the world to which we have access, can be summarized in a probability distribution 
$\mathbb{Q}$ on $\Sigma'$. Then the predicted expectation of $f$ given this information is 
$\int_{\Omega}\mathcal{E}(f)\mbox{d}\mathbb{Q}$. 
We will extend this now to quantum systems and measurements.

The guiding example is that of an $n$ level atom described by the algebra $\BC:=M_n$ undergoing 
a transformation given by a completely positive unit preserving map $T:\BC\to \BC$ 
with the following Kraus decomposition $T(X)=\sum_{i\in\Omega}V_i^*XV_i$. 
The elements of $\Omega$ can be seen as the possible measurement outcomes. 
For any initial state $\rho$ of $\BC$ and measurement result $i\in\Omega$, 
the state after the measurement is given by 
 \begin{equation*}
 \rho_i=V_i\rho V_i^*/\mbox{Tr}(V_i\rho V_i^*),
 \end{equation*}
and the probability distribution of the outcomes is $p=\sum_{i\in\Omega} p_i\delta_i$ where 
$\delta_i$ is the atomic measure at $i$, and $p_i=\mbox{Tr}(V_i\rho V_i^*)$, which 
without loss of generality can be assumed to be strictly positive. 
We represent the measurement by an instrument, that is the completely positive map with the 
following action on states
 \begin{equation}\label{eq.instrument} 
 \MC:\ M_n^*\to M_n^* \tens\ell^1(\Omega):~\rho\mapsto \sum_{i\in\Omega}\rho_i\tens p_i \delta_i.
 \end{equation} 
Let $X\in\BC$ be an observable of the system. Its expectation after the measurement, given that 
the result $i\in\Omega$ has been obtained is $\mbox{Tr}(\rho_i X)$. The function
 \begin{equation*}  
 \EC(X):\ \Omega\to \BB{C}:~  i\mapsto \mbox{Tr}(\rho_i X) 
 \end{equation*}
is the {\it conditional expectation} of $X$ onto $\ell^\infty(\Omega)$. 
If $q=\sum q_i\delta_i$ is a probability distribution on $\Omega$ then $\sum q_i\EC(X)(i)$ 
represents the expectation of $X$ on a statistical ensemble for which the distribution 
of the measurement outcomes is $q$.
We extend the conditional expectation to the linear map 
 \begin{equation*}  
 \EC:\ \BC\tens\ell^\infty(\Omega)\to\ell^\infty(\Omega)\subset \BC\tens\ell^\infty(\Omega) 
 \end{equation*} 
such that for any element $A:i\mapsto A_i$ in 
$\BC\tens\ell^\infty(\Omega)\cong \ell^\infty(\Omega\to\BC)$ we have
 \begin{equation*}  
 \EC(A):\ i\mapsto \mbox{Tr}(\rho_i A_i). 
 \end{equation*}
This map has the following obvious properties: it is idempotent and has norm one. 
Moreover, it is the unique linear map with these properties preserving the state 
$\MC(\rho)$ on $\BC\tens\ell^\infty(\Omega)$. For this reason we will call $\EC$, 
the conditional expectation with respect to $\MC(\rho)$. 
Its dual can be seen as an extension of probability distribitions 
$q\in \ell^1(\Omega)$ to states on $\BC\tens\ell^\infty(\Omega)$
 \begin{equation*} 
 \EC^*:\ q\mapsto \sum_{i\in\Omega}\rho_i\tens q_i\delta_i.
 \end{equation*}
Thus while the measurement \eqref{eq.instrument} provides a state  $\MC(\rho)$ on  
$\BC\tens\ell^\infty(\Omega)$, the conditional expectation with respect to $\MC(\rho)$ 
extends probability distributions $q\in\ell^1(\Omega)$ of outcomes, to states on 
$\BC\tens\ell^\infty(\Omega)$, and in particular on $\BC$ which represents the state after the 
measurement given the outcomes distribution $q$.

With this example in mind we pass to a more general setup which will be needed in deriving the 
stochastic Schr\"odinger equations. 
Let $\AC$ be a unital $^*$-algebra of bounded operators on a Hilbert space $\BB{H}$ whose 
selfadjoint elements represent the observables of a quantum system. 
It is natural from the physical point of view to assume that $\AC$ is strongly closed, i.e.\ 
if $\{A_n\}_{n\geq 0}$ is a sequence of operators in $\AC$ such that 
$\|A_n\psi\|\to\|A\psi\|$ for any vector $\psi$ in $\BB{H}$ and a fixed bounded operator $A$, 
then $A\in\AC$. From the mathematical point of view this leads to the rich theory of 
von Neumann algebras inspired initially by quantum mechanics, but can as well be seen 
as the generalization of probability theory to the non-commutative world of quantum mechanics. 
Indeed, the building blocks of quantum systems are matrix algebras, while probability spaces 
can be encoded into their {\it commutative} algebra of bounded random variables 
$L^\infty(\Omega, \Sigma, \BB{P})$ which appeared already in the example above. 
A state is described by a density matrix in the first case or a probability distribution 
in the second, in general it is a positive normalized linear functional $\psi:\AC\to\mathbb{C}$ 
which is continuous with respect to the weak*-topology, the natural topolgy on a von Neumann 
algebra seen as the dual of a Banach space \cite{KaR}.   
\begin{de}
Let $\BC$ be a von Neumann subalgebra of a von Neumann algebra $\AC$ of operators on a (separable)
Hilbert space $\BB{H}$. 
A \emph{conditional expectation} of $\AC$ onto $\BC$
is a linear surjective map $\EC:\ \AC \to \BC$, such that:
  \begin{enumerate}
  \item $\EC^2 = \EC$\ \ ($\EC$ is idempotent), 
  \item $\forall_{A \in \AC}:\ \| \EC(A)\| \le \| A\|$\ \ ($\EC$ is normcontractive).
  \end{enumerate}
\end{de}
In \cite{Tom} it has been shown that the conditions 
$1$ and $2$ are equivalent to $\EC$ being an identity preserving, completely positive map, 
and satisfying the \emph{module property}
 \begin{equation}\label{eq.moduleproperty}
 \EC(B_1AB_2) = 
  B_1\EC(A)B_2, \qquad\mbox{for all} ~B_1, B_2 \in \BC, ~\mbox{and}~ A \in \AC, 
 \end{equation}
generalizing a similar property of conditional expectations in 
classical probability theory (cf.\ \cite{Wil}).                                                                                     

In analogy to the classical case we are particularly 
interested in the conditional expectation which leaves a given state $\rho$ on 
$\AC$ invariant, i.e.\ $\rho\circ\EC=\rho$. 
However such a map does not always exist, but if it exists then it is unique \cite{Tak} and 
will be denoted $\EC_\rho$. 
Using $\EC_\rho$ we can extend states $\sigma$ on $\BC$ to states 
$\sigma \circ \EC_\rho$ of $\AC$ which should be interpreted as the updated state of $\AC$ 
after receiving the information (for instance through a measurement) that the subsystem 
$\BC$ is in the state $\sigma$ (cf.\ \cite{Kum3}).                     \\
In the remainder of this section we will construct the conditional expectation $\EC_\rho$
from a von Neumann algebra $\AC$ onto its \emph{center} 
$\CC := \{C \in \AC;\ AC = CA \mbox{\ for all\ } A \in \AC\}$ 
leaving a given state $\rho$ on $\AC$ invariant. The center $\CC$ is a commutative
von Neumann algebra and is therefore isomorphic to some $L^\infty(\Omega, \Sigma, \mathbb{P})$. 
In our guiding example the center of $\BC\tens\ell^\infty(\Omega)$ is $\ell^\infty(\Omega)$. 
Later on (see section \ref{QSC}) this role will be played by the 
commutative algebra of the observed process with $\Omega$ the space of all paths of 
measurement records.                               \\
\begin{stel}\label{condexp}
There exists a unique conditional expectation $\EC_\rho: \AC \to \CC$ which 
leaves the state $\rho$ on $\AC$ invariant.
\end{stel}
\begin{proof}
The proof is based on the central decomposition of $\AC$ \cite{KaR}. In our guiding example, 
$\BC\tens\ell^\infty(\Omega)$ is isomorphic to $\oplus_{i\in\Omega}\BC_i$ where 
the $\BC_i$'s are copies of $\BC$. In general we can identify the center $\CC$ with some 
$L^\infty(\Omega, \Sigma, \mathbb{P})$ where $\mathbb{P}$ corresponds to the restriction of 
$\rho$ to $\CC$. We will ignore for simplicity all issues related with measurability in the 
following constructions. 
The Hilbert space $\BB{H}$ has a direct integral representation 
$\BB{H}=\int^\oplus_\Omega \BB{H}_\omega\mathbb{P}(\mbox{d}\omega)$ in the sense that there exists 
a family of Hilbert spaces $\{\BB{H}_\omega\}_{\omega\in\Omega}$ and for any 
$\psi\in\BB{H}$ there exists a map $\omega \mapsto \psi_\omega \in \BB{H}_\omega$ such that 
 \begin{equation*}
 \langle \psi,\phi\rangle = \int_\Omega \langle \psi_\omega, \phi_\omega\rangle 
 \BB{P}(d\omega).
 \end{equation*}
The von Neumann algebra $\AC$ has a 
{\it central decomposition} $\AC= \int^\oplus_\Omega \AC_\omega \BB{P}(d\omega)$ in the sense 
that there exists a family $\{\AC_\omega\}_{\omega\in\Omega}$ of von Neumann algebras 
with trivial center, or factors, and for any $A\in\AC$ there is a map 
$\omega \mapsto A_\omega \in \AC_\omega$ such that $(A\psi)_\omega=A_\omega\psi_\omega$ 
for all $\psi\in\BB{H}$ and $\mathbb{P}$-almost all $\omega\in\Omega$. 
The state $\rho$ on $\AC$ has a decomposition in states $\rho_\omega$ on $\AC_\omega$ 
such that for any $A\in\AC$ its expectation is obtained by integrating with 
respect to $\mathbb{P}$ the expectations of its components $A_\omega$:
 \begin{equation}\label{eq.statepreserving}
 \rho(A) =  \int_\Omega \rho_\omega(A_\omega)\BB{P}(d\omega).
 \end{equation}
The map $\EC_\rho:\ \AC \to \CC$ defined by 
 \begin{equation*}
 \EC_\rho(A):\ \omega\mapsto\rho_\omega(A_\omega) 
 \end{equation*}
for all $A\in\AC$ is the desired conditional expectation. One can easily verify that this 
map is linear, identity preserving, 
completely positive (as a positive map onto a commutative von Neumann algebra), and has the 
module property. Thus, $\EC_\rho$ is a conditional expectation and leaves the state $\rho$ 
invariant by \ref{eq.statepreserving}. Uniqueness follows from \cite{Tak}.
\end{proof}

It is helpful to think of the state $\rho$ and an arbitrary operator $A$ as maps 
$\rho_\bullet:~\omega\mapsto \rho_\omega$, and respectively $A_\bullet:~\omega\mapsto A_\omega$. 
The conditional expectation $\EC_\rho(A)$ is the function 
$ \rho_\bullet(A_\bullet):~\omega\mapsto \rho_\omega(A_\omega)$.

\section{The dilation}\label{dilation}

Let $\BC$ be the observable algebra of a given quantum system on the Hilbert space $\BB{H}$. 
In the case of resonance fluorescence $\BC$ will be all $2 \times 2$ matrices $M_2$, 
the algebra of observables for the $2$-level atom. 
The irreversible evolution of the system in the Heisenberg picture is given by the 
norm continuous semigroup $\{T_t\}_{t\ge 0}$ of completely positive maps 
$T_t:\ \BC \to \BC$. By Lindblad's theorem 
\cite{Lin} we have $T_t = \exp(tL)$ where the generator $L:\ \BC \to \BC$ has the following action 
  \begin{equation}\label{eq.lindblad}
  L(X) = i[H,X] + \sum_{j=1}^k V_j^*XV_j - \frac{1}{2}\{V_j^*V_j, X\},
  \end{equation} 
where $H$ and the $V_j$'s are fixed elements of $\BC$, $H$ being selfadjoint.                          \\
We can see the irreversible evolution as stemming from a {\it reversible} 
evolution of the system $\BC$ coupled to an environment, 
which will be the electromagnetic field. 
We model a channel in the field by the bosonic or symmetric Fock space over the 
Hilbert space $L^2(\BB{R})$ of square integrable wave functions on the real line, i.e.\ 
 \begin{equation*}
 \FC := \BB{C} \oplus \bigoplus_{n=1}^\infty L^2(\BB{R})^{\ten_sn}.
 \end{equation*} 
The algebra generated by the field observables on $\FC$  contains all bounded operators and 
we denote it by $\WC$. For the dilation we will need $k$ independent copies of this algebra 
$\WC^{\ten k}$.                         \\
The free evolution of the field is given by the unitary group $S_t$, 
the second quantization of the {\it left} shift $s(t)$ on $L^2(\BB{R})$ , 
i.e. $s(t):f\mapsto f(\cdot +t)$. 
In the Heisenberg picture the evolution on $\WC$ is 
 \begin{equation*}
 W\mapsto S_t^*WS_t:=\mbox{Ad}[S_t](W).
 \end{equation*} 
The atom and field together form a closed quantum system, thus their joint evolution is 
given by a one-parameter group $\{\hat{T}_t\}_{t \in \BB{R}}$ 
of $*$-automorphisms on $\BC \ten \WC^{\ten k}$:
 \begin{equation*}
 X\mapsto \hat{U}_t^*X\hat{U}_t:=\mbox{Ad}[\hat{U}_t](X).
 \end{equation*}  
The group $\hat{U}_t$ is a perturbation of the free evolution without interaction. 
We describe this perturbation by the family of unitaries  
$U_t:=S^{\ten k}_{-t}\hat{U}_t$ for all $t\in\BB{R}$ satisfying the {\it cocycle} identity 
 \begin{equation*}
 U_{t+s} = S_{-s}^{\ten k} U_t S_{s}^{\ten k} U_s,\qquad \mbox{for all}~t,s\in\BB{R}. 
 \end{equation*}  
The direct connection between the reduced evolution of the atom given by \eqref{eq.lindblad} and 
the cocycle $U_t$ is one of the important results of quantum stochastic calculus \cite{HuP} 
which makes the object of Section \ref{QSC}. 
For the moment we only mention that in the Markov limit, 
$U_t$ is the solution of the stochastic differential equation \cite{HuP}, \cite{Par}, \cite{Mey}
  \begin{equation}\label{HuP1}
  dU_t = \{V_j dA_{j}^*(t) - V_{j}^*dA_{j}(t) -(iH + \frac{1}{2} V^*_jV_j)dt\}U_t, \qquad  U_0 
  = \mathbf{1},
  \end{equation}
where the repeated index $j$ is meant to be summed over. 
% The solution of \eqref{HuP1} forms a \emph{cocycle} with respect to the shift, i.e.\ 
% for all $t,s \in \BB{R}: U_{t+s} = S_{-s}^{\ten k} U_t S_{s}^{\ten k} U_s$. 
% The joint \emph{reversible} evolution of the system $\BC$ and the $k$ channels in 
% the field is given by the following one-parameter group $\{\hat{T}_t\}_{t \in \BB{R}}$ 
% of $*$-automorphisms on $\BC \ten \WC^{\ten k}$:
%   \begin{equation}\label{That}
%    \hat{T}_t(X) = \left\{ \begin{array}{ll}
%   U_t^{-1}S_{-t}^{\ten k}XS_t^{\ten k}U_t & \mbox{\ \ \ if \ } t \ge 0 \\
%   S_{-t}^{\ten k}U_{-t}X U^{-1}_{-t}S_t^{\ten k} & \mbox{\ \ \ if \ } t < 0 
%   \end{array}\right. ,
%   \end{equation} 
% for all $X \in \BC \ten \WC^{\ten k}$.
% It can be shown that if the cocycle satisfies equation \eqref{HuP1} we have constructed a 
% so-called \emph{quantum Markov dilation} of the dynamical system $(\BC, \{T_t\}_{t \ge 0})$. 
The quantum Markov dilation can be summarized by the following diagram 
(see \cite{Kum1}, \cite{Kum2}):
  \begin{equation}\label{dildiag}\begin{CD}
     \BC @>T_t>> \BC              \\
     @V{\Id \ten \mathbf{1}^{\ten k}}VV        @AA{\Id \ten \phi^{\ten k}}A      \\
     \BC\ten\WC^{\ten k} @>\hat{T}_t>> \BC\ten \WC^{\ten k}           \\
  \end{CD}\end{equation}
i.e.\ for all $X \in \BC:\ T_t(X) = \big(\Id \ten \phi^{\ten k}\big)
\big(\hat{T}_t(X\ten \mathbf{1}^{\ten k})\big)$, where $\phi$ is the vacuum  state on $\WC$, 
and $\mathbf{1}$ is the identity operator in $\WC$. Any dilation of the semigroup $T_t$ with 
Bose fields is unitarily equivalent with the above one under certain minimality requirements.  
The diagram can also be read in the Schr\"odinger picture if we reverse the arrows: 
start with a state $\rho$ of the system $\BC$ in the upper right hand corner, 
then this state undergoes the following sequence of maps:
  \begin{equation*}
  \rho \mapsto 
%\rho\big(\mbox{Id}\ten\phi^{\ten k}(\cdot)\big) = 
  \rho \ten \phi^{\ten k} \mapsto 
  (\rho \ten \phi^{\ten k})\circ\hat{T}_t =\hat{T}_{t*}(\rho \ten \phi^{\ten k})
  \mapsto 
  \mbox{Tr}_{\FC^{\ten k}}\big(\hat{T}_{t*}(\rho \ten \phi^{\ten k})\big).
  \end{equation*} 
This means that at $t=0$, the atom in state $\rho$ is coupled to the $k$ channels in the 
vacuum state, and after $t$ seconds of unitary evolution we take the 
partial trace taken over the $k$ channels.                                                                           

We would now like to introduce the measurement process. It turns out that this can be best 
described in the {\it interaction picture}, where we let the shift part of 
$\hat{U}_t=S^{\tens k}_tU_t$ act on the observables while the cocycle part acts on the states:
 \begin{equation}\label{eq.stateinteraction}
 \rho^t(X) := \rho\ten\phi^{\ten k}(U_t^*XU_t)
 \end{equation} 
for all $X \in \BC\ten \WC^{\ten k}$. It is well known that for the Bose field for arbitrary 
time $t$ we can split the noise algebra as a tensor product
 \begin{equation*}
 \WC = \WC_{0)} \ten \WC_{[0,t)} \ten \WC_{[t}
 \end{equation*} 
with each term being the algebra generated by those fields over test functions with support in 
the corresponding subspace of $L^2(\BB{R})$:
 \begin{equation*}
L^2(\BB{R}) = 
L^2\big((-\infty, 0)\big)\oplus L^2\big([0, t)\big) \oplus L^2\big([t, \infty)\big).
\end{equation*} 
Such a continous tensor product structure is called a {\it filtration} and it is essential in the 
development of quantum stochastic calculus reviewed in Section \ref{QSC}. 
The observables which we measure in an arbitrary time interval $[0,t)$ form a commuting 
family of selfadjoint operators $\{Y_s\}_{0\leq s\leq t}$ 
whose spectral projections belong to the  middle part of the tensor product $\WC_{[0,t)}$. 
In the Davies process $Y_s=\Lambda(s)$, i.e.\ the number operator up to time $s$, while  in 
the homodyne case $Y_s=X_\phi(s)$. 
Notice that the part $\WC_{0)}$ will not play any significant role as it corresponds to ``what 
happened before we started our experiment''.

Let $\CC_t$ be the commutative von Neumann generated by 
the observed process up to time $t$, $\{Y_s\}_{0 \le s \le t}$ $(t \ge 0)$, 
seen as a subalgebra of $\BC\ten \WC^{\ten k}$. By a theorem on von Neumann algebras, 
$\CC_t$ is equal to the double commutant of the observed process up 
to time $t$: $\CC_t =\{Y_s;\ 0\le s \le t\}''$, with the \emph{commutant} $\SC'$ of a 
subset $\SC$ of $\BC\ten \WC^{\ten k}$ being defined by 
$\SC' := \{X \in \BC\ten \WC^{\ten k};\ XS = SX\ \forall S\in \SC \}$. 
The algebras $\{\CC_t\}_{t\geq 0}$ form a growing family, that is 
$\CC_s\subset\CC_t$ for all $s\leq t$. Thus we can define the 
inductive limit $\CC_{\infty}:=\lim_{t\to\infty}\CC_t$, which is the smallest von Neumann 
algebra containing all $\CC_{t}$. On the other hand for each $t\geq 0$ we have a 
state on $\CC_t$ given by the restriction of the state $\rho^t$ of the whole system 
defined by \eqref{eq.stateinteraction}. We will show now that that the states 
$\rho^t$ for different times ``agree with each other''.
\begin{stel} 
On the commutative algebra $\CC_{\infty}$ there exists a unique 
state $\rho^\infty$ which coincides with $\rho^t$ when restricted to $\CC_t\subset\CC_{\infty}$, 
for all $t\geq 0$. In particular there exists a measure space 
$(\Omega,\Sigma,\mathbb{P}_\rho)$ such that $(\CC_\infty,\rho^\infty)$ is isomorphic with 
$L^\infty(\Omega,\Sigma,\mathbb{P}_\rho)$ and a growing family $\{\Sigma_t\}_{t\geq 0}$ of 
$\sigma$-subalgebras of $\Sigma$ such that  
$(\CC_t,\rho^t)\cong L^\infty(\Omega,\Sigma_t,\mathbb{P}_\rho)$.
\end{stel}

\begin{proof} 
In the following we will drop the extensive notation of tensoring identity operators 
when representing operators in $\WC_{[s,t)}$ for all $s,t\in\BB{R}$. 
Let $X\in \CC_s$, in particular $X\in\WC_{[0,s)}^{\ten k}$. 
By \eqref{HuP1}, $U_t\in \BC\ten\WC_{[0,t)}^{\ten k}$, because the coefficients of the stochastic 
differential equation lie in $\BC \ten \WC_{[0,t)}^{\ten k}$. This implies that 
$S_{-s}^{\ten k}U_tS_s^{\ten k}\in \BC\ten \WC_{[s,t+s)}^{\ten k}$. Using the tensor product structure 
of $\WC^{\ten k}$, we see that $\WC_{[0,s)}^{\ten k}$ and $\BC\ten\WC_{[s,t+s)}^{\ten k}$ commute, 
and in particular $X$ commutes with $S_{-s}^{\ten k}U_tS_s^{\ten k}$. Then 
\begin{eqnarray}\label{consistent}
  \rho^{t+s}(X) &=& \rho^{0}(U_{t+s}^* X U_{t+s}) = 
  \rho^0(U_s^*(S_{-s}^{\ten k}U_tS_s^{\ten k})^*  X S_{-s}^{\ten k}U_tS_s^{\ten k}U_s)  
  \nonumber\\
 &=&\rho^0\big(U_s^* X U_s\big) =  \rho^s(X). 
  \end{eqnarray}
This implies that the limit state $\rho^\infty$ on $\CC_\infty$ with the desired 
properties exists, in analogy to the Kolmogorov extension theorem for probability measures. 
As seen in the previous section, $(\CC_\infty,\rho^\infty)$ is isomorphic to 
$L^\infty(\Omega, \Sigma, \BB{P}_\rho)$ for some 
probability space $(\Omega, \Sigma, \BB{P}_\rho)$. 
The subalgebras $(\CC_t,\rho^t)$ are isomorphic to 
$L^\infty(\Omega,\Sigma_t,\mathbb{P}_\rho)$ for some growing family $\{\Sigma_t\}_{t\geq 0}$ of 
$\sigma$-subalgebras of $\Sigma$. 
% Every set $E$ in $\Sigma^s$ for some $s \ge 0$ corresponds to the 
% projection in $L^\infty(\Omega^s, \Sigma^s, \BB{P}^s_\rho)$ given by the characteristic 
% function $\chi_E$ of the set $E$. 
% For all $t \ge 0$ we find, using the cocycle property for the family $\{U_r\}_{r \in \BB{R}}$:
 %  \begin{equation}\begin{split}\label{consistent}
%   &\BB{P}^{t+s}_\rho[E] = \rho^{t+s}(I_\BC \ten \chi_E \ten I_{\WC_{[s}}) = 
%   \rho^0\big(U_{t+s}^*(I_\BC \ten \chi_E \ten I_{\WC_{[s}})U_{t+s}\big) =   \\
%   &\rho^0\big(U_s^*(S_{-s}^{\ten k}U_tS_s^{\ten k})^*
%   (I_\BC \ten \chi_E \ten I_{\WC_{[s}})S_{-s}^{\ten k}U_tS_s^{\ten k}U_s\big) =  \\
%   &\rho^0\big(U_s^*(I_\BC \ten \chi_E \ten I_{\WC_{[s}})U_s\big) =  \BB{P}_\rho^s[E], 
%   \end{split}\end{equation}
% where we used that for all $r\in \BB{R}:\ U_r \in \BC \ten \WC^{\ten k}_{[0,r)}$, 
% i.e.\  $S_{-s}^{\ten k}U_tS_s^{\ten k} \in \BC \ten \WC^{\ten k}_{[s,s+t)}$ commutes with 
% $I_\BC \ten \chi_E \ten I_{\WC_{[s}}$. In the following we will drop the extensive notation 
% tensoring identity operators to operators in $\WC_{[0,r)}$. Equation \eqref{consistent} shows
% that the family of probability measures $\{\BB{P}^t_\rho\}_{t \ge 0}$ is consistent, hence
% by Kolmogorov's extension theorem it extends to a single probability measure $\BB{P}_\rho$ on the
% sigma field $\Sigma^\infty$, generated by the sigma fields $\{\Sigma^t\}_{t\ge 0}$. 
% This means the observed process $\{Y_s\}_{s\ge 0}$ is a family of random variables 
% on one single probability space $(\Omega^\infty, \Sigma^\infty, \BB{P}_\rho)$.                   
\end{proof} 

\noindent\textbf{Remark.}                                     
From spectral theory it follows that the measure space $(\Omega^t,\Sigma_t)$ coincides with the 
joint spectrum of $\{Y_s\}_{s\leq t}$, i.e.\ $\Omega^t$ is the set of all paths of the 
process up to time $t$. For the example of the counting process this means that $\Omega^t$ is 
the Guichardet space of the interval $[0,t)$, which is the set of all sets of instants 
representing a "click" of the photon counter, i.e.\ it is the set of all paths of the 
counting process.              

We define now $\ \AC_t := \CC_t'$ for all $t\geq 0$ , 
i.e.\ $\AC_t$ is the commutant of $\CC_t$, then 
$\CC_t$ is the center of the von Neumann algebra $\AC_t$. Notice that the observable algebra 
of the atom $\BC$ is contained in $\AC_t$. By Theorem \ref{condexp} we can construct a family 
of conditional expectations $\{\EC_{\rho^t}^t:\ \AC_t \to \CC_t\}_{t\ge 0}$. 
For each $t$, $\EC_{\rho^t}^t$ depends on the state of the ``world'' at that moment 
$\rho^t$, keeping this in mind we will simply denote it by $\EC^t$. An important property
of $\EC^t$ is that $\rho^\infty \circ \EC^t = \rho^t \circ \EC^t = \rho^t$, since the range
of $\EC^t$ is $\CC_t$ and $\EC^t$ leaves $\rho^t$ invariant. 

For an element $X \in \AC_t$, $\EC^t(X)$ is an element in $\CC_t$, i.e.\ a 
function on $\Omega_t$. Its value in a point $\omega\in\Omega_t$, i.e.\ an outcomes 
record up to time $t$, is the expectation value of $X$ given the observed path 
$\omega$ after $t$ time units. 
We will use the notation $\EC^t(X):=\rho^t_\bullet(X_\bullet)$ defined in the end of Section 
\ref{conditional expectation} to emphasise the fact that this is a function on $\Omega_t$. 
When restricted to $\BC\tens \CC_t$ the conditional expectation is precisely of the type 
discussed in our guiding example in Section \ref{conditional expectation}.\\
There exists no conditional expectation from $\BC \ten \WC$ onto $\CC_t$ since 
performing the measurement has \emph{demolished} the information about observables that 
do not commute with the observed process \cite{Bel}. 
We call $\AC_t$ the algebra of observables that are 
\emph{not demolished} \cite{Bel} by observing the process $\{Y_s\}_{0\le s\le t}$. 
This means that performing the experiment and ignoring the outcomes gives the same time 
evolution on $\AC_t$ as when no measurement was done. 

From classical probability it follows that 
for all $t\ge 0$ there exists a unique conditional expectation $\BB{E}^t_\rho:\ \CC_\infty \to \CC_t$
that leaves the state $\rho^\infty$ invariant, i.e.\ $\rho^\infty \circ \BB{E}^t_\rho = \rho^\infty$.
These conditional expectations have the {\it tower property}, 
i.e.\ $\BB{E}^s_\rho\circ\BB{E}^t_\rho = \BB{E}^s_\rho$ for all $t \ge s \ge 0$, which is often very
useful in calculations. $\BB{E}^0_\rho$ is the expectation with respect to $\BB{P}_\rho$, and will
simply be denoted $\BB{E}_\rho$. Note that the tower property for $s=0$ is exactly the invariance of 
the state $\rho^\infty (= \BB{E}_\rho)$.

\section{Quantum stochastic calculus}\label{QSC}

In this section we briefly discuss the quantum stochastic calculus developed by Hudson and 
Parthasarathy \cite{HuP}. For a detailed treatment of the subject we refer to 
\cite{Par} and \cite{Mey}. 
Let $\FC(\HC)$ denote the symmetric (or bosonic) Fock space over the one particle space 
$\HC :=\mathbb{C}^k\ten L^2(\BB{R}_+) = L^2(\{1,2,\ldots k\}\times \BB{R}_+)$. 
% i.e.\ $\FC(\HC) = \BB{C}\oplus \bigoplus_{n=1}^\infty \HC^{\ten_s n}$. 
% Here $l^2(\{1,2,\ldots k\})$ is 
% the space of all $k$-tuples of complex numbers with the obvious inner product and 
% $L^2(\BB{R})$ is the space of all quadratically integrable wave functions on $\BB{R}_+$.     
The space $\mathbb{C}^k$ describes the $k$ channels we identified in the 
electromagnetic field. As in the previous section we denote the algebra of bounded operators 
on the one channel Fock space $\FC (\BB{R}_+)$ by $\WC$, and on the $k$ channels 
$\FC (\HC)$ by $\WC^{\tens k}$. \\
For every $f \in \HC$ we define the \emph{exponential vector} $e(f) \in \FC(\HC)$ 
in the following way:
\begin{equation*}
e(f) := 1 \oplus \bigoplus_{n=1}^\infty \frac{1}{\sqrt{n!}}f^{\ten n}, 
\end{equation*}
which differs from the coherent vector by a normalization factor. The inner products of
two exponential vectors $e(f)$ and $e(g)$ is 
$\langle e(f),e(g)\rangle=\mbox{exp}(\langle f,g\rangle)$. 
Note that the span of all exponential vectors, denoted $\DC$, forms a dense subspace of 
$\FC(\HC)$. 
Let $f_j$ be the $j$'th component of $f \in \HC$ for $j=1,2,\ldots,k$. 
%$x \in \BB{R}_+:\ f_j(x) := \langle \delta_j,  f(x)\rangle_{l^2(\{1,2,\ldots, k\})}$. 
The annihilation operator $A_j(t)$, creation operator $A^*_j (t)$ and number operator 
$\Lambda_{ij} (t)$ are defined on the domain $\DC$ by
  \begin{equation*}\begin{split}
  &A_j(t)e(f) = \langle \chi_{[0,t]},\, f_j \rangle e(f) = \int_0^t f_j(s)ds~ e(f)\\
  &\big\langle e(g),\, A^*_j(t) e(f) \big\rangle = 
   \langle g_j,\, \chi_{[0,t]} \rangle \big\langle e(g),\, e(f) \big\rangle = 
   \int_0^t \overline{g}_j(s)ds ~ \mbox{exp}(\langle f,g\rangle)           \\
  &\big\langle e(g),\, \Lambda_{ij}(t) e(f) \big\rangle =
  \langle g_i,\, \chi_{[0,t]}f_j\rangle\big\langle e(g),\, e(f) \big\rangle = 
  \int_0^t \overline{g}_i(s) f_j(s)ds ~\mbox{exp}(\langle f,g\rangle) .
  \end{split}\end{equation*} 
The operator $\Lambda_{ii}(t)$ is the usual counting operator for the $i$'th channel. 
Let us write $L^2(\BB{R^+})$ as direct sum $L^2([0,t]) \oplus L^2([t,\infty])$, 
then $\FC(L^2(\BB{R}_+))$ is unitarily equivalent with 
$\FC(L^2([0,t])\ten \FC(L^2[t,\infty))$ through the identification 
$e(f) \cong e(f_{t]}) \ten e(f_{[t})$, with $f_{t]} = f\chi_{[0,t]}$ and 
$f_{[t} = f\chi_{[t,\infty)}$. We will also use the notation $f_{[s,t]}$ for $f\chi_{[s,t]}$ 
and omit the tensor product signs between exponential vectors. 
The same procedure can be carried out for all the $k$ channels.   
                      
Let $M_t$ be one of the processes $A_j(t), A^*_j(t)$ or $\Lambda_{ij}(t)$.
The following factorisability property \cite{HuP}, \cite{Par} makes  
the definition of stochastic integration against $M_t$ possible
  \begin{equation*}
  (M_t - M_s)e(f) = e(f_{s]})\big\{(M_t-M_s) e(f_{[s,t]})\big\}e(f_{[t}), 
  \end{equation*} 
with $(M_t-M_s) e(f_{[s,t]}) \in \FC\big(\mathbb{C}^k\tens L^2([s,t])\big)$. 
We firstly define the stochastic integral for the so called  
{\it simple} operator processes  with values in the 
atom and noise algebra $\BC \ten \WC^{\tens k}$ where $\BC:=M_n$

\begin{de} 
Let $\{L_s\}_{0 \le s \le t}$ be an adapted 
(i.e.\ $L_s \in \BC \ten \WC_{s]}$ for all $0 \le s \le t$) simple 
process with respect to the partition $\{s_0=0, s_1,\dots, s_p= t\}$ in the sense that 
$L_s = L_{s_j}$ whenever $s_j \le s < s_{j+1}$. Then the stochastic integral of 
$L$ with respect to $M$ on $\BB{C}^n \ten \DC$ is given by \cite{HuP}, \cite{Par}:  
  \begin{equation*}
  \int_0^t L_s dM_s  ~f e(u) := 
\sum_{j=0}^{p-1} \big(L_{s_j}fe(u_{s_j]})\big)\big((M_{s_{j+1}}- M_{s_{j}})
   e(u_{[s_j, s_{j+1}]})\big)e(u_{[s_{j+1}}).
  \end{equation*}
\end{de}
By the usual approximation by simple processes we can extend the definition of the 
stochastic integral 
to a large class of {stochastically integrable processes} \cite{HuP}, \cite{Par}. 
We simplify our notation by writing $dX_t = L_tdM_t$ for 
$X_t = X_0 + \int_0^t L_sdM_s$. Note that the definition of the stochastic integral 
implies that the increments 
$dM_s$ lie in the future, i.e.\ $dM_s \in \WC_{[s}$. Another consequence of the definition 
of the stochastic 
integral is that its expectation with respect to the vacuum state $\phi$ is always $0$ due to 
the fact that the increments $dA_j, dA^*_j,d\Lambda_{ij}$ have zero expectation values in the 
vacuum. This will often simplify calculations of expectations, our strategy being that of 
trying to bring these increments to act on the 
vacuum state thus eliminating a large number of differentials. 

The following theorem of Hudson and Parthasarathy extends the It\^o rule 
of classical probability theory.\\
\begin{stel}\label{Itorule}\textbf{(Quantum It\^o rule \cite{HuP}, \cite{Par})}
Let $M_1$ and $M_2$ be one of the processes $A_j, A^*_j$ or $\Lambda_{ij}$. Then 
$M_1M_2$ is an adapted process satisfying the relation:
  \begin{equation*}
  dM_1M_2 = M_1dM_2 + M_2dM_1 + dM_1dM_2,
  \end{equation*}
where $dM_1dM_2$ is given by the quantum It\^o table:
\begin{center}
{\large \begin{tabular} {l|lll}
$dM_1\backslash dM_2$ & $dA^*_i$ & $d\Lambda_{ij}$ & $dA_i$ \\
\hline 
$dA^*_k$ & $0$ & $0$ & $0$ \\
$d\Lambda_{kl}$ & $\delta_{li}dA^*_k$ & $\delta_{li}d\Lambda_{kj}$ & $0$  \\
$dA_k$ & $\delta_{ki}dt$ & $\delta_{ki}dA_j$ & $0$ 
\end{tabular} }
\end{center}
\end{stel}
\noindent\textbf{Notation.} The quantum It\^o rule will be used for 
calculating differentials of products of It\^o integrals. 
Let $\{Z_i\}_{i=1,\dots, p}$ be It\^o integrals, then 
 \begin{equation*} 
 d (Z_1Z_2\dots Z_p)= \sum_{\substack{\nu\subset\{1,\dots, p\} \\ \nu \neq \emptyset}}[\nu]
 \end{equation*}
where the sum runs over all {\it non-empty} subsets of $\{1,\dots, p\}$ and for any 
$\nu=\{i_1,\dots i_k\}$, the term $[\nu]$ is the contribution 
to $d (Z_1Z_2\dots Z_p)$ coming from differentiating only the terms with indices 
in the set $\{i_1,\dots i_k\}$ and preserving the order of the factors in the product. 
For example the differential $d(Z_1Z_2Z_3)$ contains terms of the type 
$[2]=Z_1(dZ_2)Z_3$, $[13]=(dZ_1)Z_2(dZ_3)$, and $[123]=(dZ_1)(dZ_2)(dZ_3)$.

Let $V_j$ for $j = 1,2,\ldots,k$, and $H$ be operators in $\BC$ with $H$ is selfadjoint. 
Let $S$ be a unitary operator on $\BB{C}^n \ten l^2(\{1,2,\ldots, k\})$ with 
$S_{ij}=\langle i, S j \rangle\in\BC$ the ``matrix elements'' in the basis 
$\{|i>:i=1,\dots, k\}$ of $\mathbb{C}^k$. Then there exists a unique unitary 
solution for the following quantum stochastic differential equation \cite{HuP}, \cite{Par}
  \begin{equation}\label{HuP2}
  dU_t = \big\{V_jdA_j^*(t) + (S_{ij}- \delta_{ij})d\Lambda_{ij}(t) - 
  V^*_iS_{ij}dA_j(t) -(iH + \frac{1}{2} V_j^*V_j)dt\big\}U_t, 
  \end{equation}
with initial condition $U_0 = \I$, where again repeated indices have been summed. 
Equation \eqref{HuP1}, providing the cocycle of unitaries perturbing the free evolution of the 
electromagnetic field is an example of such an equation. 
The terms $d\Lambda_{ij}$ in equation $\eqref{HuP2}$ describe 
direct scattering between the channels in the electromagnetic field \cite{BaL}.
We have omitted this effect for the sake of simplicity, i.e.\ we always take 
$S_{ij} = \delta_{ij}$.  \\ 
We can now check the claim made in Section \ref{dilation} that the dilation diagram 
\ref{dildiag} commutes. It is easy to see that following the lower part of the diagram 
defines a semigroup on $\BC$. We have to show it is generated by $L$. 
For all $X \in \BC$ we have 
 \begin{equation*}
 d~\Id\ten\phi^k\big(\hat{T}_t(X\ten \I^{\ten k})\big)
 = \Id\ten\phi^k(d~U_t^* X\ten \I^{\ten k} U_t). 
 \end{equation*}
Using the It\^o rules we obtain
 \begin{equation*}
 d~U_t^* X\ten \I^{\ten k} U_t = (dU_t^*)X\ten \I^{\ten k} U_t + U_t^*X\ten \I^{\ten k}dU_t + 
 (dU_t^*)X\ten \I^{\ten k} dU_t. 
 \end{equation*}
With the aid of the It\^o table we can evaluate these terms. 
We are only interested in the $dt$-terms since the expectation with respect to the 
vacuum kills the other terms. 
Then we obtain: $d~\Id\ten\phi^k(U_t^* X\ten \I^{\ten k} U_t) = 
\Id\ten\phi^k(U_t^* L(X)\ten \I^{\ten k} U_t)dt$, proving the claim. 

Now we return to the example of resonance fluorescence. Suppose the laser is off,
then we have spontaneous decay of the $2$-level atom into the field which is in the vacuum state. 
For future convenience we already distinguish
a \emph{forward} and a \emph{side} channel in the field, the Liouvillian  is then given by
  \begin{equation*}
  L(X) = i[H, X] + \sum_{\sigma = f,s} V_\sigma^* X V_\sigma - 
  \frac{1}{2}\{V^*_\sigma V_\sigma, X\},  
  \end{equation*}  
where 
  \begin{equation*}
  V = \begin{pmatrix} 0 & 0 \\ 1 & 0 \end{pmatrix},\ \ \ V_f = \kappa_fV,\ \ \ V_s = 
  \kappa_sV,\ \ \ 
  |\kappa_f|^2 + |\kappa_s|^2 = 1, 
  \end{equation*} 
with $|\kappa_f|^2$ and $|\kappa_s|^2$ the decay rates into the forward and side channel 
respectively.   \\
The dilation of the quantum dynamical system $(M_2, \{T_t =\exp(tL)\}_{t \ge 0})$, 
is now given by the closed system 
$(M_2\ten\WC_f\ten\WC_s, \{\hat{T}_t\}_{t\in \BB{R}})$ with unitary 
cocycle given by
  \begin{equation*}
  dU^{sd}_t = \{V_fdA^*_f(t)-V^*_fdA_f(t) + V_sdA^*_s(t) - V_s^*dA_s(t) - 
  (iH+\frac{1}{2} V^*V)dt\}U^{sd}_t, \ \ 
  U^{sd}_0 = \I,
  \end{equation*}  
where the superscript $sd$ reminds us of the fact that the laser is off, 
i.e.\ we are considering spontaneous decay. We can summarize this in the following 
dilation diagram
  \begin{equation*}\begin{CD}
     \BC @>T_t = \exp(tL)>> \BC              \\
     @V{\Id \ten \I \ten \I}VV        @AA{\Id \ten \phi\ten\phi}A      \\
     \BC\ten\WC_f\ten\WC_s @>\hat{T}^{sd}_t = \mbox{Ad}[\hat{U}^{sd}_t]>> \BC\ten \WC_f\ten\WC_s           \\
  \end{CD}\end{equation*}
where $\hat{U}^{sd}_t$ is given by $S_t\ten S_t U^{sd}_t$ for $t \ge 0$.                            \\
We change this setting by introducing a laser on the forward channel, i.e.\ 
the forward channel is now in a coherent state (see \ref{eq.coherent}) 
$\gamma_h := \langle\psi(h),\,\cdot\,\psi(h)\rangle$ for some $h \in L^2(\BB{R}_+)$. 
This leads to the following dilation diagram
  \begin{equation}\label{laserdiagram}\begin{CD}
     \BC @>T^h_t>> \BC              \\
     @V{\Id \ten \I \ten \I}VV        @AA{\Id \ten \gamma_h\ten\phi}A      \\
     \BC\ten\WC_f\ten\WC_s @>\hat{T}^{sd}_t = \mbox{Ad}[\hat{U}^{sd}_t]>> \BC\ten \WC_f\ten\WC_s           \\
  \end{CD}\end{equation}
i.e.\ the evolution on $\BC$ has changed and it is in general {\it not} a semigroup. 
Denote by $W(h)$ the unitary \emph{Weyl} or 
\emph{displacement operator} defined on $\DC$ by: 
$W(h)\psi(f) = \exp(-2i\mbox{Im}\langle h, f\rangle)\psi(f+h)$. 
% This definition leads to a linear isometric operator $W(g)$ on the dense domain $\DC$ which 
% uniquely extends to the whole of $\FC$. 
Note that $W(h)\phi = W(h)\psi(0) = \psi(h)$, so that we can write
  \begin{equation*}\begin{split}
  &T^h_t(X) = \mbox{Id} \ten \gamma_h \ten \phi({U^{sd}_t}^*X\ten \I\ten \I U^{sd}_t) = 
  \mbox{Id} \ten \phi \ten \phi\big(W_f(h)^* {U^{sd}_t}^*X\ten \I\ten \I U^{sd}_t W_f(h)\big) =  \\
  &\mbox{Id} \ten \phi \ten \phi\big(W_f(h_{t]})^* {U^{sd}_t}^*X\ten \I\ten \I U^{sd}_t W_f(h_{t]})\big),
  \end{split}\end{equation*}                    
where $h_{t]} := h\chi_{[0,t]}$ and $W_f(h):=\I\tens W(h)\tens\I$. 
Defining $U_t := U^{sd}_tW_f(h_{t]})$, together with the stochastic differential 
equation for $W_f(h_{t]})$ \cite{Par}
  \begin{equation*}
  dW_f(h_{t]}) = \{h(t) dA^*_f(t) -\overline{h}(t)dA_f(t) - 
  \frac{1}{2}|h(t)|^2dt\}W_f(h_{t]}),\ \ \ \ W_f(h_0) = \I, 
  \end{equation*}
and the It\^o rules leads to the following quantum stochastic differential equation for $U_t$:
  \begin{eqnarray*}
  dU_t=& \big\{(V_f+h(t))dA^*_f(t) -(V_f^* + \overline{h}(t))dA_f(t)\ + 
  V_sdA^*_s(t) - V^*_sdA_s(t)\ -  \\
  &-\big(iH+\frac{1}{2}(|h(t)|^2 + V^*V + 2h(t)V_f^*)\big)dt~\big\} U_t,\ \ \ U_0 = \I.
  \end{eqnarray*}
Define $\tilde{V}_f := V_f + h(t)$, $\tilde{V}_s := V_s$ and 
$\tilde{H} := H + i\frac{1}{2}(\overline{h}(t)V_f - h(t)V_f^*)$ then this reads
  \begin{equation}\label{cocyclelaser}
dU_t =\ \sum_{\sigma = f,s}\big\{ \tilde{V}_\sigma dA^*_\sigma(t) - 
  \tilde{V}^*_\sigma dA_\sigma 
  - \frac{1}{2}(i \tilde{H} + \tilde{V}^*_\sigma\tilde{V}_\sigma)dt\big\}U_t, \ \ \ U_0 = \I.
  \end{equation}
The time dependent generator of the dissipative evolution in the presence of 
the laser on the forward channel is 
  \begin{equation}\label{timedepgenerator}
  L(X) = i[\tilde{H}, X] + \sum_{\sigma = f,s}\tilde{V}^*_\sigma X\tilde{V}_\sigma - 
       \frac{1}{2}\{\tilde{V}^*_\sigma\tilde{V}_\sigma, X\}.
  \end{equation}
Therefore the diagram for resonance fluorescence \eqref{laserdiagram} is equivalent to
  \begin{equation*}\label{laserII}\begin{CD}
     \BC @>T^h_t>> \BC              \\
     @V{\Id \ten \I \ten \I}VV        @AA{\Id \ten \phi\ten\phi}A      \\
     \BC\ten\WC_f\ten\WC_s @>\hat{T}_t = \mbox{Ad}[\hat{U}_t] >> \BC\ten W_f\ten\WC_s           \\
  \end{CD}\end{equation*}  
where $\hat{U}_t$ is given by $S_t\ten S_t U_t$ for $t \ge 0$.
For $h(t) = -i \Omega/\kappa_f$, we find the master equation for resonance 
fluorescence \eqref{Master}. From now on we will no longer suppress the oscillations of 
the laser, i.e.\ we take $h(t) = -i \exp(i\omega t)\Omega/\kappa_f$. Then we find
  \begin{equation*}
  L(X) = i[H, X] - 
  i\frac{\Omega}{2}[e^{-i\omega t}V+e^{i\omega t}V^*, X] - \frac{1}{2}\{V^*V,X\} + V^* X V,
  \end{equation*}
note that the laser is resonant when $\omega = \omega_0$.

\section{Belavkin's stochastic Schr\"odinger equations}\label{Belavkinslemma}

Now we are ready to derive a stochastic differential equation for the process $\EC^t(X)$. 
In the next section we will see that this equation leads to the 
stochastic Schr\"odinger equations \eqref{Belcount} and \eqref{Belhomodyne}, that we 
already encountered in Sections \ref{Daviesprocess} and \ref{Homodynedetection}. 

\begin{de}\label{defMt}
Let $X$ be an element of $\BC:=M_n$. Define the process $\{M^X_t\}_{t\ge0}$ in the algebra 
$\CC_\infty \cong L^\infty(\Omega, \Sigma, \BB{P}_\rho)$, generated by the observed process 
$\{Y_t\}_{t\ge 0}$ (see Section \ref{dilation}) by
  \begin{equation*}
  M^X_t := \EC^t(X) - \EC^0(X) - \int_0^t \EC^r\big(L(X)\big)dr, 
  \end{equation*}
where $L:\ \BC \to \BC$ is the Liouvillian. 
In the following we suppress the superscript $X$ in $M^X_t$ 
to simplify our notation.  
\end{de}

Note that from the above definition it is clear that $M_t$ is an element of $\CC_t$ for all 
$t\ge 0$. The following theorem first appeared (in a more general form and with a different proof) 
in \cite{Bel} and is at the heart of quantum filtering theory. We prove it using the properties 
of conditional expectations. For simplicity we have restricted to observing a process in the 
field $\WC^{\ten k}$. The theory can be extended to processes that are in $\BC\ten \WC^{\ten k}$,
transforming it into a more interesting filtering theory. For the stochastic Schr\"odinger equations
arising in quantum optics our approach is general enough. 

\begin{stel}\label{BelavkinI}
The process $\{M_t\}_{t\ge 0}$ of definition \ref{defMt} is a martingale with respect to 
the filtration $\{\Sigma_t\}_{t\ge 0}$ of $\Omega$ and the measure $\BB{P}_\rho$, i.e.\ 
for all $t \ge s \ge 0$ we have: $\BB{E}_\rho^s(M_t) = M_s$. 
\end{stel} 
\begin{proof}
From the module property of the conditional expectation it follows that $\BB{E}_\rho^s(M_t) = M_s$ 
for $t \ge s \ge 0$ is equivalent to $\BB{E}_\rho^s(M_t -M_s) = 0$ for $t \ge s \ge 0$. This means 
we have to prove for all $t \ge s \ge 0$ and $E \in \Sigma_s$: 
  \begin{equation*}
  \int_E\BB{E}_\rho^s(M_t-M_s)(\omega)\BB{P_\rho}(d\omega) = 0,
  \end{equation*}
which, by the tower property, is equivalent to
  \begin{equation}\label{martprop}
  \int_E (M_t-M_s)(\omega)\BB{P_\rho}(d\omega) = 0,
  \end{equation}
i.e.\  $\BB{E}_\rho\big(\chi_E(M_t-M_s)\big) = 0$. 
Now using Definition \ref{defMt} and again the module property of the conditional 
expectation we find,
writing $E$ also for the projection corresponding to $\chi_E$
  \begin{equation*}\begin{split}
  \BB{E}_\rho\big(\chi_E(M_t-M_s)\big) &= 
  \rho^\infty\Big(\EC^t(X\ten E) - \EC^s(X\ten E) - 
   \int_s^t\EC^r\big(L(X)\ten E\big)dr\Big)                                 \\
  &= \rho^t(X \ten E) - \rho^s(X \ten E) - \int_s^t \rho^r\big(L(X)\ten E\big)dr. 
  \end{split}\end{equation*}
This means we have to prove: $d\rho^t(X \ten E) - \rho^t\big(L(X) \ten E\big)dt = 0$,
for all $t \ge s$. Note that $\rho^t(X\ten E) = \rho^0(U^*_t X \ten E U_t) = 
\rho \ten \phi^{\ten k}(U^*_t X \ten E U_t)$. Therefore $d\rho^t(X\ten E) = 
\rho \ten \phi^{\ten k}\big(d(U^*_t X\ten E U_t)\big)$. We will use the notation
below Theorem \ref{Itorule} with $Z_1 = U^*_t$ and $Z_2 = X \ten E U_t$. Using
the quantum It\^o table and the fact that only the $dt$ terms 
survive after taking a vacuum expectation, we find:
  \begin{equation*}\begin{split}
  d\rho^0(U^*_tX \ten EU_t) &= \rho^0\big([1]\big) + \rho^0\big([2]\big) + 
     \rho^0\big([12]\big), \ \ \mbox{where} \\
  \rho^0\big([1]\big) + \rho^0\big([2]\big) & = 
     \rho^0\big(U^*_t(i[H,X]\ten E - \frac{1}{2}\{V_j^*V_j, X\}\ten E) U_t\big)dt \\
  \rho^0\big([12]\big) & = \rho^0\big(U^*_t(V^*_jXV_j)\ten EU_t\big)dt. 
  \end{split}\end{equation*}
This means $d\rho^t(X\ten E) = \rho^t\big(L(X)\ten E\big)dt$, for all $t \ge s$, 
proving the theorem.
\end{proof}
Note that in the proof of the above theorem we have used that the projection $E \in \CC_s$ commutes 
with the increments $dA_{j}(s),\, dA^*_{j}(s),\, ds$ and with the processes in front of the 
increments in equation \eqref{HuP1}, i.e.\ $V_j,\, V_j^*,\, V_j^*V_j$ and $H$. 
If the theory is extended to a more 
general filtering theory \cite{Bel}, then these requirements become real restrictions on 
the process $\{Y_t\}_{t \ge 0}$. If they are satisfied the observed process 
$\{Y_t\}_{t \ge 0}$ is said to be \emph{self non demolition} \cite{Bel}.          \\
Definition \ref{defMt} implies the following stochastic differential equation 
for the process $\EC^t(X)$
  \begin{equation}\label{Belav1}
  d\EC^t(X) = \EC^t\big(L(X)\big)dt + dM_t, 
  \end{equation}
called the \emph{Belavkin equation}. The only thing that remains to be done is linking the 
increment $dM_t$ to the increment of the observed process $dY_t$.                        \\
Let us assume that the observed process $\{Y_t\}_{t \ge 0}$ satisfies a quantum stochastic 
differential equation
  \begin{equation*}
  dY_t =  \alpha_j(t) dA^*_{j}(t) + \beta_{ij}(t) d\Lambda_{ij}(t) +  \alpha_j^*(t)dA_{j}(t)  + 
         \delta(t) dt,
  \end{equation*}
for some adapted stochastically integrable processes 
$\alpha_j, \beta_{ij}$, and $\delta$, such that 
$\alpha_j(t), \beta_{ij}(t),\delta(t)\in\WC^{\ten k}_{t]}$ for all $t\geq 0$, and 
$\beta_{ij}^*=\beta_{ji},\, \delta=\delta^*$ since $Y_t$ is selfadjoint. Furthermore, since 
the observed process $\{Y_t\}_{t\ge 0}$ is commutative, 
we have $[dY_t, Y_s] = 0$ for all $s \le t$, which 
leads to
  \begin{equation*}\begin{split}
  &[\alpha_j(t), Y_s]dA^*_j(t) + [\beta_{ij}(t), Y_s]d\Lambda_{ij}(t) + [\alpha^*_j(t), Y_s]dA_j(t)
  + [\delta(t), Y_s]dt = 0  \ \  \Rightarrow \\
  & [\alpha_j(t), Y_s] = 0, \ \ [\beta_{ij}(t), Y_s] = 0, \ \ [\alpha^*_j(t), Y_s] = 0, \ \ 
  [\delta(t), Y_s] = 0,
  \end{split}\end{equation*}
i.e.\ $\alpha_j(t), \beta_{ij}(t), \alpha^*_j(t), \delta(t) \in \AC_t$.
This enables us to define a process $\tilde{Y}_t$ by 
  \begin{equation}\label{dYtilde}\begin{split}
  d\tilde{Y}_t =&\ \Big( \alpha_j(t)dA^*_j(t) - \EC^t\big(V_j^*\alpha_j(t)\big)dt \Big)\ + 
            \Big(\beta_{ij}(t) d\Lambda_{ij}(t) - \EC^t\big(V_i^*\beta_{ij}(t)V_j\big)dt\Big)\ + \\
                &\Big(\alpha_j^*(t)dA_j(t) - \EC^t\big(\alpha_j^*(t)V_j\big)dt\Big) , 
           \ \ \ \ \tilde{Y}_0 = 0,  
  \end{split}\end{equation} 
i.e.\ we have the following splitting of $Y_t$:
  \begin{equation}\label{DoobMeyer}
  Y_t = Y_0 + \tilde{Y}_t + \int_0^t \Big(\EC^s\big(V_j^*\alpha_j(s)\big) + 
  \EC^s\big(V^*_i\beta_{ij}(s)V_j\big) 
   + \EC^s\big(\alpha_j^*(s)V_j\big) + \delta(s)\Big)ds, 
  \end{equation}
which in view of the following theorem is the semi-martingale splitting of $Y_t$. The process 
$\tilde{Y}_t$ is called the \emph{innovating martingale} of the observed process $Y_t$.

\begin{stel}\label{BelavkinII}
The process $\{\tilde{Y}_t\}_{t \ge 0}$ is a martingale with respect to 
the filtration $\{\Sigma_t\}_{t\ge 0}$ of $\Omega$ and the measure $\BB{P}_\rho$, i.e.\ 
for all $t \ge s \ge 0$ we have: $\BB{E}_\rho^s(\tilde{Y}_t) = \tilde{Y}_s$.
\end{stel}
\begin{proof}
We need to prove that for all $t \ge s \ge 0:\ \BB{E}^s_\rho(\tilde{Y}_t -\tilde{Y}_s) = 0$. 
This means we have to prove for all $t \ge s \ge 0$ and $E \in \Sigma_s$:
  \begin{equation*}\begin{split}
  &\int_E \BB{E}^s_\rho(\tilde{Y}_t -\tilde{Y}_s)(\omega) \BB{P}_\rho(d\omega) = 0 \iff 
   \int_E (\tilde{Y}_t -\tilde{Y}_s)(\omega) \BB{P}_\rho(d\omega) = 0 \iff \\
  &\BB{E}_\rho \Bigg(Y_tE - Y_sE - \int_s^t \Big(\EC^r\big(V_j^*\alpha_j(r)\big)E + 
  \EC^r\big(V^*_i\beta_{ij}(r)V_j\big)E 
   + \EC^r\big(\alpha_j^*(r)V_j\big)E + \delta(r)E\Big)dr\Bigg) = 0 \iff \\
  &\rho^t(Y_tE) - \rho^s(Y_sE) =  \int_s^t \rho^r\Big(\EC^r\big(V_j^*\alpha_j(r)\big)E + 
  \EC^r\big(V^*_i\beta_{ij}(r)V_j\big)E 
   + \EC^r\big(\alpha_j^*(r)V_j\big)E + \delta(r)E\Big)dr.
  \end{split}\end{equation*}
For $t=s$ this is okay, so it remains to be shown that for all 
$t \ge s \ge 0$ and $E \in \Sigma_s$:
  \begin{equation*}\begin{split}
  &d\rho^t(Y_tE) = \rho^t\Big(\EC^t\big(V_j^*\alpha_j(t)\big)E +
  \EC^t\big(V^*_i\beta_{ij}(t)V_j\big)E + \EC^t\big(\alpha_j^*(t)V_j\big)E + 
  \delta(t)E\Big)dt\ \iff \\
  &d\rho^0(U_t^*Y_tEU_t) = \rho^t\Big(\EC^t\big(V_j^*\alpha_j(t)\big)E +
  \EC^t\big(V^*_i\beta_{ij}(t)V_j\big)E + \EC^t\big(\alpha_j^*(t)V_j\big)E + \delta(t)E\Big)dt.
  \end{split}\end{equation*}
We define: $Z_1(t) := U^*_t,\ Z_2(t) := Y_tE$ and $Z_3(t) := U_t$ then we find, using the
notation below Theorem \ref{Itorule}: 
$d\rho^0(U_t^*Y_tEU_t) = \rho^0([1] + [2] + [3] + [12] + [13] + 
[23] + [123])$. 
Remember $\rho^0 = \rho \ten \phi^{\ten k}$, i.e.\ we are only interested in the $dt$ terms, 
since the vacuum kills all other terms. The terms $[1], [3]$ and $[13]$ together make up 
the usual Lindblad term and since $L(\I) = 0$ we do not have to consider them.                  \\
Furthermore, term [2] contributes $U^*_t\delta(t)EU_tdt$, term [12] contributes 
$U^*_tV_j^*\alpha_j(t)EU_tdt$,
term $[23]$ contributes $U^*_t\alpha^*_j(t)V_jEU_tdt$ and term $[123]$ contributes 
$U^*_tV_i^*\beta_{ij}(t)V_jU_tdt$, therefore we get
  \begin{equation*}\begin{split}
  &d\rho^0(U_t^*Y_tEU_t) = \rho^0\big(U^*_t\alpha^*_j(t)V_jEU_t + U^*_tV_i^*\beta_{ij}(t)V_jU_t +
  U^*_tV_j^*\alpha_j(t)EU_t + U^*_t\delta(t)EU_t\big)dt = \\
  &\rho^t\big(\alpha^*_j(t)V_jE + V_i^*\beta_{ij}(t)V_j + V_j^*\alpha_j(t)E + 
     \delta(t)E\big)dt = \\
  &\rho^t\Big(\EC^t\big(V_j^*\alpha_j(t)\big)E +
  \EC^t\big(V^*_i\beta_{ij}(t)V_j\big)E + \EC^t\big(\alpha_j^*(t)V_j\big)E + \delta(t)E\Big)dt,
  \end{split}\end{equation*}
proving the theorem.      
\end{proof} 

\noindent\textbf{Remark.} In the probability literature an adapted process which can be written 
as the sum of a martingale and a finite variation process is called a semimartingale \cite{RoW}. 
The Theorems \ref{BelavkinI} and \ref{BelavkinII} show that $M_t$ and $Y_t$ are semimartingales.

We now represent the martingale $M_t$ from Definition \ref{defMt} as an integral over the 
innovating martingale (cf.\ \cite{Kal}) by
\begin{equation}\label{martingalerep}
dM_t = \eta_t d\tilde{Y}_t
\end{equation} 
for some stochastically integrable process $\eta_t$, which 
together with equation \eqref{DoobMeyer} provides the link between $dM_t$ and $dY_t$. 
We are left with the problem of determining 
$\eta_t$, which we will carry out in the next section for the examples 
of Section \ref{Daviesprocess} and \ref{Homodynedetection}. 
Here we just give the recipe for finding $\eta_t$. 

\textbf{Recipe.} Define for all integrable adapted 
processes $b_t$ and $c_t$ a process $B_t$ in $\CC_\infty$ by
  \begin{equation}\label{eq.Bt}
  dB_t = b_t d\tilde{Y_t} + c_t dt.
  \end{equation}
These processes form a dense subalgebra of $\CC_\infty$. 
Now determine $\eta_t$ from the fact that 
$\EC^t$ leaves $\rho^t$ invariant \cite{Bel}, i.e.\ for all $B_t$
  \begin{equation*}
  \rho^t\big(\EC^t(B_tX)\big) = \rho^t(B_tX). 
  \end{equation*}
From this it follows that for all $B_t$ 
 \begin{equation}\label{eq.for.k}
  d\rho^0\big(U_t^*B_t(\EC^t(X) - X)U_t\big) = 0. 
  \end{equation}
We evaluate the differential $d\big(U_t^*B_t(\EC^t(X) - X)U_t\big)$ using the 
quantum It\^o rules. Since $\rho^0 = \rho \ten \phi^{\ten k}$ we can restrict to the 
$dt$ terms, since the others die on the vacuum. We will use the notation below 
Theorem \ref{Itorule} with $Z_1(t) = U_t^*,\ Z_2(t) = B_t, 
Z_3(t) = \EC^t(X)-X$ and $Z_4(t) = U_t$. The following lemma simplifies the 
calculation considerably.  

%Since the equation \ref{eq.for.k} are valid for all $b_t,c_t$ and therefore also for all 
%$B_t$, we have to set the coefficients in front of these terms to zero. 
%It is easy to see that the terms containing $c_t$ add up to zero. 
%This is also true for the terms with $B_t$
\begin{lem} The sum of all terms in which $Z_2$ is not differentiated has zero expectation:  
$\rho^0 ([1]+[3]+[4]+[13]+[14]+[34]+[134])=0$.
\end{lem}

\begin{proof}
The $dt$ terms of $[3]$ are $U^*_tB_t\EC^t\big(L(X)\big)U_tdt$ and 
$-U^*_tB_t\eta_t\big(\EC^t(V^*_j\alpha_j) + \EC^t(V^*_i\beta_{ij}V_j) + 
\EC^t(\alpha^*_jV_j)\big)U_tdt$. Using the fact that $\EC^t$ leaves $\rho^t$ invariant
we see that the term $U^*_tB_t\EC^t\big(L(X)\big)U_tdt$ cancels 
against the $dt$ terms of  $[1], [4]$ and $[14]$, which make up the Lindblad generator $L$
with a minus sign. The other term of $[3]$ is cancelled in expectation against the $dt$ 
terms of $[13], [34]$ and $[134]$, since 
  \begin{equation*}\begin{split}
  &\rho^0([13]) = \rho^t(B_t\eta_t V^*_j\alpha_j)dt =  
    \rho^t\big(\EC^t(B_t\eta_t V^*_j\alpha_j)\big)dt = 
    \rho^t\big(B_t\eta_t \EC^t(V^*_j\alpha_j)\big)dt\\
  &\rho^0([34]) = \rho^t(B_t\eta_t \alpha^*_jV_j)dt =  
    \rho^t\big(\EC^t(B_t\eta_t \alpha^*_jV_j)\big)dt =
    \rho^t\big(B_t\eta_t\EC^t(\alpha^*_jV_j)\big)dt \\
  &\rho^0([134]) = \rho^t(B_t\eta_t V^*_i\beta_{ij}V_j)dt =
     \rho^t\big(\EC^t(B_t\eta_t V^*_i\beta_{ij}V_j)\big)dt = 
     \rho^t\big(B_t\eta_t \EC^t(V^*_i\beta_{ij}V_j)\big)dt.
  \end{split}\end{equation*}  
\end{proof}

Using equation \eqref{dYtilde}, the fact that $\EC^t$ leaves $\rho^t$ invariant and the 
module property, we find that the term $[2]$ has expectation 
zero as well 
 \begin{equation*}\begin{split} 
  &\rho^0([2]) = \rho^t\left(b_td\tilde{Y}_t(\EC^t(X)-X)\right) =  
  -\rho^t\left(b_t\EC^t(V_j^*\alpha_j(t)+\alpha_j^*(t)V_j+V_i^*\beta_{ij}V_j)(\EC^t(X)-X)\right) dt = \\
  &-\rho^t\left(b_t\EC^t(V_j^*\alpha_j(t)+\alpha_j^*(t)V_j+V_i^*\beta_{ij}V_j)\EC^t(\EC^t(X)-X)\right)dt = 
  0.
 \end{split}\end{equation*}
Thus, only the terms containing no $B_t$ nor $c_t$ can contribute non-trivially. This leads to an  
equation allowing us to obtain an expression for $\eta_t$ by solving
  \begin{equation}
  \rho^0([12]+[23]+[24]+[123]+[124]+[234]+[1234])=0.
  \end{equation}
Although this can be carried out in full generality, we will provide the solution only for our 
main examples, the photon counting and homodyne detection experiments for  
a resonance fluorescence setup, in the next section.

\section{Examples}

We now return to the example considered in Section \ref{Daviesprocess}. 
We were considering a $2$-level
atom in interaction with the electromagnetic field. The interaction was given by a cocycle $U_t$ 
satisfying equation \eqref{cocyclelaser}. The observed process is
the number operator in the side channel, i.e.\ $Y_t = \Lambda_{ss}(t)$. 
Therefore $d\tilde{Y}_t = d\Lambda_{ss}(t) - \EC^t(V_s^*V_s)dt$. 
Recall now the notation $Z_1(t) = U_t^*,\ Z_2(t) = B_t, Z_3(t) = \EC^t(X)-X$ and $Z_4(t) = U_t$,
their differentials are given by
  \begin{equation*}\begin{split}
  dU^*_t =&\ U^*_t \sum_{\sigma = f,s} \big\{ \tilde{V}^*_\sigma dA_\sigma(t) - 
\tilde{V}_\sigma dA^*_\sigma(t) 
  - \frac{1}{2}(-i \tilde{H} + \tilde{V}^*_\sigma\tilde{V}_\sigma)dt\big\}                         \\
  dB_t =&\ b_td\Lambda_{ss}(t) + \big(c_t - b_t\EC^t(V_s^*V_s)\big)dt                        \\
  d(\EC^t(X)-X) =&\ \eta_t d\Lambda_{ss}(t) + \Big(\EC^t\big(L(X)\big) - 
\eta_t\EC^t(V_s^*V_s)\Big)dt \\
  dU_t =&\ \sum_{\sigma = f,s}\big\{ \tilde{V}_\sigma dA^*_\sigma(t) - 
\tilde{V}^*_\sigma dA_\sigma(t) 
  - \frac{1}{2}(i \tilde{H} + \tilde{V}^*_\sigma\tilde{V}_\sigma)dt\big\}U_t.
  \end{split}\end{equation*}
Following the recipe of the previous section we now only have to determine the $dt$ terms of
$[12],[23],[24],$ $[124], [123], [124]$ and $[1234]$. All of these terms are zero in expectation 
with respect to $\rho^0$, except for $[124]$ and $[1234]$ 
  \begin{equation*}\begin{split}
  \rho^0\big([124]\big) &= \rho^0\Big(U_t^* b_tV^*_s\big(\EC^t(X) -X\big)V_s U_t\Big)dt  \\
  \rho^0\big([1234]\big) &= \rho^0\big(U_t^*b_t\eta_tV_s^*V_s U_t\big)dt.
  \end{split}\end{equation*}
For all $b_t$ the sum of these terms has to be  $0$ in expectation, i.e.
  \begin{equation*}\begin{split}
  &\forall b_t:\ \rho^t\Bigg(b_t\Big(V^*_s\big(\EC^t(X) -X\big)V_s +     
  \eta_tV_s^*V_s\Big) \Bigg)dt = 0 \iff \\
  &\forall b_t:\ \rho^t\Bigg(\EC^t\Bigg(b_t\Big(V^*_s\big(\EC^t(X) -X\big)V_s +     
  \eta_tV_s^*V_s\Big) \Bigg)\Bigg)dt = 0 \iff \\
  &\forall b_t:\ \rho^t\Bigg(b_t\Big(\EC^t(X)\EC^t(V^*_sV_s) -\EC^t(V^*_sXV_s) +     
  \eta_t\EC^t(V_s^*V_s)\Big) \Bigg)dt = 0 \iff \\
  &\eta_t = \frac{\EC^t(V^*_sXV_s)}{\EC^t(V_s^*V_s)}- \EC^t(X).
  \end{split}\end{equation*} 
Substituting the expressions for $\eta_t$ and $\tilde{Y}_t$ into equation \eqref{Belav1} 
we obtain the 
Belavkin equation for photon counting in the side channel
  \begin{equation}\label{BelavcountHeis}
  d\EC^t(X) = \EC^t\big(L(X)\big)dt + 
  \Big(\frac{\EC^t(V^*_sXV_s)}{\EC^t(V_s^*V_s)}
  - \EC^t(X)\Big)\big(d\Lambda_{ss}(t) - \EC^t(V_s^*V_s)dt\big).
  \end{equation}
Now recall that $\EC^t(X) = \rho^t_\bullet (X_\bullet)$, i.e.\ it is the function 
$\Omega_t \to \BB{C}:\ 
\omega \mapsto \rho^t_\omega(X_\omega)$. For all $X \in \BC = M_2$, the $M_2$ valued function 
$X_\bullet$
is the constant function $\omega \mapsto X$. Therefore for all $X$ in $\BC$, 
the Belavkin equation \eqref{BelavcountHeis} is equivalent to
  \begin{equation*}
  d\rho_\bullet^t(X) = \rho_\bullet^t\big(L(X)\big)dt + 
  \Big(\frac{\rho_\bullet^t(V^*_sXV_s)}{\rho_\bullet^t(V_s^*V_s)}
  - \rho_\bullet^t(X)\Big)\big(d\Lambda_{ss}(t) - \rho_\bullet^t(V_s^*V_s)dt\big),
  \end{equation*}
which is equivalent to the Belavkin equation of Section \ref{Daviesprocess}, equation 
\eqref{Belcount}. In simulating the above equation we can take for $Y_t = \Lambda_{ss}(t)$
the unique jump process with independent jumps and rate $\rho^t_\bullet(V_s^*V_s)$, since
$\Lambda_{ss}(t) - \int_0^t \rho^r_\bullet(V_s^*V_s)dr$ has to be a martingale.

Let us now turn to the homodyne detection scheme which we already discussed in Section 
\ref{Homodynedetection}.
The observed process is now $Y_t = X_\phi(t) = A^*_s(f_t) + A_s(f_t)$ (see Section 
\ref{Homodynedetection} for the definition of $f_t$). 
This means the 
innovating martingale $\tilde{Y}_t$ satisfies 
$d\tilde{Y}_t = e^{i\phi_t}dA^*_s(t) + e^{-i\phi_t}dA_s(t) - \EC^t(e^{i\phi_t}V^*_s + e^{-i\phi_t}V_s)dt$,
where $\phi_t = \phi_0 + \omega_{lo}t$ with $\omega_{lo}$ the frequency of the local oscillator.
Therefore we find different differentials for $B_t$ and $\EC^t(X)-X$ than we had in the photon counting
case
  \begin{equation*}\begin{split}
  dB_t =&\ b_t\big(e^{i\phi_t}dA^*_s(t) + e^{-i\phi_t}dA_s(t)\big) + 
     \big(c_t - b_t\EC^t(e^{i\phi_t}V^*_s + e^{-i\phi_t}V_s)\big)dt                        \\
  d(\EC^t(X)-X) =&\ \eta_t\big(e^{i\phi_t}dA^*_s(t) + e^{-i\phi_t}dA_s(t)\big)  + 
      \Big(\EC^t\big(L(X)\big) - \eta_t\EC^t(e^{i\phi_t}V^*_s + e^{-i\phi_t}V_s)\Big)dt \\
  \end{split}\end{equation*}
Following the recipe of the previous section we now only have to determine the $dt$ terms of
$[12],[23],$ $[24], [124], [123], [124]$ and $[1234]$. All of these terms are zero in expectation 
with respect to $\rho^0$, except for $[12], [23]$ and $[24]$ 
  \begin{equation*}\begin{split}
  \rho^0\big([12]\big) &= \rho^0\Big(U_t^* e^{i\phi_t}V^*_sb_t\big(\EC^t(X)-X\big)U_t\Big)dt\\
  \rho^0\big([23]\big) &= \rho^0(U_t^*b_t\eta_t U_t)dt\\
  \rho^0\big([24]\big) &= \rho^0\Big(U_t^*b_t\big(\EC^t(X)-X\big)e^{-i\phi_t}V_s U_t\Big)dt.
  \end{split}\end{equation*}
For all $b_t$ the sum of these terms has to be  $0$ in expectation, i.e.
  \begin{equation*}\begin{split}
  &\forall b_t:\ \rho^t\Bigg(b_t\Big(e^{i\phi_t}V^*_s\big(\EC^t(X)-X\big)+ 
  \big(\EC^t(X)-X\big)e^{-i\phi_t}V_s + \eta_t\Big)\Bigg)dt = 0 \iff \\
  &\forall b_t:\ \rho^t\Bigg(\EC^t\Bigg(b_t\Big(e^{i\phi_t}V^*_s\big(\EC^t(X)-X\big)+ 
  \big(\EC^t(X)-X\big)e^{-i\phi_t}V_s + \eta_t\Big)\Bigg)\Bigg)dt = 0 \iff \\
  &\forall b_t:\ \rho^t\Big(b_t\big( -\EC^t(e^{i\phi_t}V^*_sX  + e^{-i\phi_t}XV_s) +
    \EC^t(e^{i\phi_t}V^*_s  + e^{-i\phi_t}V_s)\EC^t(X)+ \eta_t \big)\Big)dt = 0 \iff \\
  &\eta_t = \EC^t(e^{i\phi_t}V^*_sX  + e^{-i\phi_t}XV_s) - \EC^t(e^{i\phi_t}V^*_s  + 
e^{-i\phi_t}V_s)\EC^t(X).
  \end{split}\end{equation*} 
Substituting the expressions for $\eta_t$ and $\tilde{Y}_t$ into equation \eqref{Belav1} we obtain the 
Belavkin equation for the homodyne detection scheme
  \begin{equation}\label{BelavhomoHeis}\begin{split}
  &d\EC^t(X) = \EC^t\big(L(X)\big)dt + 
  \big(\EC^t(e^{i\phi_t}V^*_sX  + e^{-i\phi_t}XV_s) - \EC^t(e^{i\phi_t}V^*_s  + 
e^{-i\phi_t}V_s)\EC^t(X)\big)\times\\
  &\times \big(e^{i\phi_t}dA^*_s(t) + e^{-i\phi_t}dA_s(t) - \EC^t(e^{i\phi_t}V^*_s + 
e^{-i\phi_t}V_s)dt\big).
  \end{split}\end{equation}
Now recall that $\EC^t(X) = \rho^t_\bullet (X_\bullet)$, i.e.\ it is the function 
$\Omega_t \to \BB{C}:\ \omega \mapsto \rho^t_\omega(X_\omega)$. For all $X \in \BC = M_2$, the $M_2$ 
valued function $X_\bullet$
is the constant function $\omega \mapsto X$. 
Therefore for all $X$ in $\BC$, the Belavkin equation \eqref{BelavhomoHeis} is equivalent to
  \begin{equation*}\begin{split}
  &d\rho_\bullet^t(X) = \rho_\bullet^t\big(L(X)\big)dt + 
  \big(\rho_\bullet^t(e^{i\phi_t}V^*_sX  + e^{-i\phi_t}XV_s) - \rho_\bullet^t(e^{i\phi_t}V^*_s  + 
  e^{-i\phi_t}V_s)\rho_\bullet^t(X)\big)\times  \\
  &\times \big(e^{i\phi_t}dA^*_s(t) + e^{-i\phi_t}dA_s(t) - 
  \rho_\bullet^t(e^{i\phi_t}V^*_s + e^{-i\phi_t}V_s)dt\big),
  \end{split}\end{equation*}
which is equivalent to the Belavkin equation of Section \ref{Homodynedetection}, equation 
\eqref{Belhomodyne}. 
Since $A^*_s(f_t) + A_s(f_t) - \int_0^t\rho_\bullet^r(e^{i\phi_r}V^*_s + e^{-i\phi_r}V_s)dr$ 
is a martingale with variance $t$ on the space of the Wiener process, it must be the Wiener process itself.

\bibliography{filtering}

\end{document}